\documentclass[aps,preprintnumbers,superscriptaddress,twocolumn,amsmath,amssymb,floatfix,pra]{revtex4} 
\pdfoutput=1
\usepackage{graphicx}
\usepackage[usenames,dvipsnames]{color}
\usepackage[colorlinks,bookmarks=false,citecolor=NavyBlue,linkcolor=Red,urlcolor=blue]{hyperref}
\usepackage[export]{adjustbox}

\newcommand\be{\begin{equation}}
\newcommand\bea{\begin{eqnarray}}
\newcommand\bes{\begin{subequations}}
\newcommand\esu{\end{subequations}}
\newcommand\ee{\end{equation}}
\newcommand\eea{\end{eqnarray}}

\newcommand{\cmmnt}[1]{}

\newcommand\ba         {\begin{eqnarray} } 
\newcommand\ea         {\end{eqnarray} }

\def\doi{http://dx.doi.org/}

\usepackage{braket}

\newcommand\ocite[1]{[\onlinecite{#1}]}
\newcommand{\dd}{{\rm d}}

\begin{document}

\title{Lack of thermalization for integrability-breaking impurities}

\author{Alvise Bastianello}
\affiliation{Institute for Theoretical Physics, University of Amsterdam, Science Park 904, 1098 XH Amsterdam, The Netherlands}

\date{\today}
\begin{abstract}
We investigate the effects of localized integrability-breaking perturbations on the large times dynamics of thermodynamic one-dimensional quantum and classical systems.
In particular, we suddenly activate an impurity which breaks the integrability of an otherwise homogeneous system.
We focus on the large times dynamics and on the thermalization properties of the impurity, which is shown to have mere perturbative effects even at infinite times, thus preventing thermalization.
This is in clear contrast with homogeneous integrability-breaking terms, which display the prethermalization paradigm and are expected to eventually cause thermalization, no matter the weakness of the integrability-breaking term.
Analytic quantitative results are obtained in the case where the bulk Hamiltonian is free and the impurity interacting.
\end{abstract}

\maketitle

\section{Introduction} 
Recent experimental advances in the cold atom's world \cite{exp1,exp2,exp3,exp4,exp5,exp6,exp7,exp8,exp9,exp10,exp11,exp12,exp13,exp14,exp15} caused an outburst of theoretical efforts aimed to understand the out-of-equilibrium properties of closed many-body quantum systems.
In particular, dimensional reduction and the extreme precision in the coupling tunability
gave access to the one-dimensional world, allowing the experimental realization of several playgrounds for theoretical physicists, such as  integrable models \cite{Korepin,smirnov,takahashi}.
Integrable systems possess infinitely many local conserved quantities, which deeply affect their out-of-equilibrium features: after an homogeneous quantum quench \cite{calabrese-cardy}, local observables relax to a steady state which is not described by the usual thermal ensemble. 
The information encrypted in the conserved degrees of freedom is retained up to infinite time and the Gibbs Ensemble urges a modification: this led to the construction of the Generalized Gibbs Ensemble (GGE) $e^{-\sum_i\beta_i \hat{\mathcal{Q}}_i}$ \cite{ggerigol,ggew1,ggew2,ggew3,ggew4,ggew5,ggew6,ggew7,ggew8,ggew9,ggew10,ViRi2016,specialissue}, where all the relevant (quasi-)local conserved charges $\hat{\mathcal{Q}}_i$ \cite{ggef1,ggef2,ggef3,ggef4, lch1,lch2,lch3,lch4,lch5,lch6,lchf1,lchf2,lchf3,lchf4,lchf5} are kept into account.
In view of the remarkable difference between non-integrable and integrable models, 
understanding the effect of integrability-breaking perturbations is a central question, from both a theoretical and experimental point of view.
In particular, what is the destiny of a system with weakly broken integrability?
The homogeneous case has been thoroughly  investigated in the last years and the so called ``prethermalization" \cite{bertini_prethermal,pre_th_1,pre_th_2,pre_th_3,pre_th_4,pre_th_5,pre_th_6,pre_th_7,pre_th_8,pre_th_9,pre_th_10,pre_th_11,pre_th_12,pre_th_13,pre_th_14,pre_th_15,Alba_Fagotti_prethermal} paradigm has been identified. 
Local observables relax in two steps: on a short time scale the system apparently reaches a non trivial GGE state, built on the integrable part of the Hamiltonian. Subsequently, a slow drift towards a final thermal ensemble is observed: such a picture found experimental confirmation \cite{exp_pre_th1,exp_pre_th2,exp_pre_th3,exp_pre_th4}.
Crucially, the magnitude of the integrability breaking term affects the time scale on which the thermal ensemble is attained \cite{bertini_prethermal}, but does not spoil the dichotomy between integrable and non-integrable systems. 
In contrast with the homogeneous case, the effect of localized integrability-breaking terms has not been systematically assessed so far.
The activation of a localized perturbation could seem a rather innocent operation, but it has tremendous consequences on a fragile property such as integrability \cite{localquenchesBertini_Fagotti,fag_non_th_def}.
More specifically, we consider at time $t<0$ an integrable Hamiltonian $\hat{H}_{\text{I}}$, the infinite system being initialized in a suitable homogeneous GGE. For $t\ge 0$, we activate a localized perturbation $\hat{V}(x)$, which we refer to as ``defect" or  ``impurity"
\be
\hat{H}=\hat{H}_\text{I}+\int_{-\Delta}^{\Delta} \dd x\, \hat{V}(x)\, .
\ee

Generalizing the modes of free systems, integrable Hamiltonians are diagonalized in terms of multi-particle states \cite{Korepin,smirnov,takahashi}. Because of the infinite set of constraints due to the conserved charges, these quasiparticles necessarily undergo only elastic pairwise scattering events. In the thermodynamic limit, homogeneous GGEs are in a one-to-one correspondence \cite{lch4} with a set of root densities \cite{takahashi}, which describe the density of quasi-particles with a given momentum $k$. For simplicity, we restrict ourselves to the case of a single root density $\rho(k)$.

Although the impurity breaks the integrability of the system as a whole, far from the defect the system is locally described by an integrable Hamiltonian, with stable quasi-particle excitations. Therefore, in the spirit of the Generalized Hydrodynamics  (GHD) \cite{GHD1,GHD2} (see also Ref. \cite{DoyonSphon17,Doyon17,GHD3,GHD6,GHD7,GHD8,GHD10,F17,DS,ID117,DDKY17,DSY17,ID217,CDV17,mazza2018,BFPC18}), at large times and far from the defect the system locally relaxes to an inhomogeneous GGE \cite{localquenchesBertini_Fagotti,localquenches4,Bas_DeL_hop,Bas_DeL_ising}.
 The latter is fully determined by an inhomogeneous root density $\rho_{x,t}(k)$, with the appealing semiclassical interpretation of a local density of particles.  Due to the ballistic spreading characteristic of integrable models, the propagating GGE only depends on the ``ray" $\zeta=x/t$: such a state has been named Local Quasi Stationary State (LQSS) \cite{GHD2}.
In contrast, the limit $t\to \infty$ with finite $x$ is known as the Non Equilibrium Stationary State (NESS) \cite{ruelle2000,bedo15,localquenchesBertini_Fagotti,fag15}.

While localized impurities have been frequently studied in a whole variety of contexts (see in particular Ref. \cite{GoMiOl18,PoAhHe18} for integrability-breaking issues), the large times dynamics in the present framework has been analyzed only in free models and CFTs \cite{localquenches1,localquenches3,localquenches4,Bas_DeL_hop,Bas_DeL_ising}, being the defect free or CFT invariant respectively (see however \cite{localquenchesBertini_Fagotti}).
Instead, we eventually consider $\hat{V}$ to be an integrability-breaking interaction.

From the semiclassical viewpoint, quasiparticles undergo non-elastic scattering events while crossing the defect's region, leading immediately to the natural central question of the present work: how does the density root of the quasiparticles emerging from the defect look like?
At large times, a finite subsystem encompassing the defect will relax to the NESS. Reasonably, the latter can be expected to be described by a GGE's density matrix and, being the Hamiltonian the only conserved charge, it appears natural to revert to thermal states.
Based on the insight gained in the homogeneous case, long transients can be imagined, but a thermal state should be eventually reached:
in this case, the quasiparticles emerging from the defect should be thermally distributed.
However, preliminary numerical results go against this natural expectation \cite{fag_non_th_def}.

In this work we study the thermalizing properties of integrability-breaking impurities. In contrast with the homogeneous case, we show how a weak integrability-breaking defect has poor mixing properties which ultimately prevent thermalization.
We focus on a free theory in the bulk, but with an interacting defect: we build a perturbative expansion of the LQSS in the strength of the interaction, which is finite at any order.
For technical reasons clarified later on, we focus on continuum models which are not suited for efficient numerical methods such as DMRG \cite{dmrg_rev}.
However, the same questions can be posed in classical models (see Ref. \cite{DeLuca_Mussardo2016,BDWY2018} for the construction of GGE and GHD in classical integrable field theories), which allow for a numerical benchmark.
We expect the same general conclusions to hold true also in truly interacting integrable models, in view of the following heuristic argument.

\section{Some heuristic considerations} 
It is widely accepted that standard time-dependent perturbation theory \cite{sakurai} is not suited to study the late-time physics of thermodynamically large homogeneous systems, making necessary to resort to other methods \cite{bertini_prethermal,Nessi_Iucci}.
This is due to secular terms that grow unbounded in time: thermalization is intrinsically a non-perturbative effect.
However, simple heuristic arguments point out the possible perturbative nature of the defect.
We semiclassically regard the initial state as a gas of quasiparticles which, for $t>0$, undergo inelastic scattering within the support of the perturbation.
In the case of a homogeneous integrability-breaking term,
a given quasiparticle takes part in a growing number of inelastic scatterings, piling up a cumulative effect which leads to secular terms.
In contrast, in the the impurity case, a traveling quasiparticle can undergo inelastic processes only on the defect's support, where it typically spends a finite amount of time. Therefore, small inelastic scatterings cannot sum up to an appreciable contribution, suggesting the perturbative nature of the impurity.
Moreover, it could be argued that in the low density limit only the few-body physics rules the dynamics on the defect. Few-body physics has weak thermalizing properties, see eg. Ref. \cite{YuOl10} for the two particles case.
Albeit physically sounding, these considerations remain on an heuristic ground and can be regarded as plausible as the previous reasoning, based on the conserved charges: in this perspective, a rigorous and well-controlled benchmark is needed.

\section{A specific model} In order to test our ideas, we consider a chain of harmonic oscillators
\be\label{HI}
\hat{H}_\text{I}=\int_{-\infty}^\infty \dd x\,\, \frac{1}{2}\Big\{\hat{\Pi}^2(x)+[\partial_x\hat{\phi}(x)]^2+m^2\hat{\phi}^2(x)\Big\}\, ,
\ee
where $\hat{\Pi}(x)$ and $\hat{\phi}(x)$ are conjugated fields $[\hat{\phi}(x),\hat{\Pi}(y)]=i\delta(x-y)$.
Being $\hat{H}_\text{I}$ free, is of course also integrable.
Generalizations to other free models, such as Galilean-invariant bosons and fermions (see the Supplementary Material (SM) \cite{suppl}), are straightforward.

The integrability-breaking potential is chosen as a function of $\hat{\phi}(x)$, i.e. $\hat{V}(x)=V(\hat{\phi}(x))$.
$\hat{H}_I$ is diagonalized in the Fourier space in terms of bosonic operators $[\hat{a}(k),\hat{a}^\dagger(q)]=\delta(k-q)$ \cite{peskin} and the modes are readily interpreted as the quasiparticles with energy $E(k)=\sqrt{k^2+m^2}$ and velocity $v(k)=\partial_k E(k)$. 
GGEs are simply gaussian ensembles in $a(k)$ \cite{gaussification1,gaussification2,gaussification3,gaussification4} (and in the field $\hat{\phi}$), with the root density being associated with the mode density $\langle a^\dagger(k)a(q)\rangle=\delta(k-p)\rho(k)$. 
Therefore, the two point function computed on a GGE $\Gamma_{t-\tau}(x-y)=\langle\hat{\phi}_t(x)\hat{\phi}_{\tau}(y)\rangle_\text{GGE}$ is
\be\label{corr}
\Gamma_t(x) =\int_{-\infty}^\infty \frac{\dd k}{2\pi}\frac{\cos(E(k)t-kx)}{E(k)}\rho(k)+\frac{\cos(kx)}{2E(k)}e^{-iE(k)t}\, .
\ee
At large times, after the defect activation and far from it, the GHD prediction states that the two point correlator has the same expression as above, provided we replace $\rho(k)\to \rho_{x,t}(k)$ \cite{GHD1,GHD2}
\be\label{inhroot}
\rho_{x,t}(k)=\rho(k)+\Theta(|v(k) |-|\zeta|)\Theta(v(k) \zeta)\delta \rho(k)\, .
\ee
Above, $\Theta$ is the Heaviside Theta function and $\zeta=x/t$. The interpretation is clear: for $t>0$ a perturbation of the initial root density $\delta\rho(k)$ ballistically propagates from the impurity, affecting only a finite interval of length $|t v(k)|$ placed on the right(left) of the defect for $v(k)>0$ ($v(k)<0$). 
The computation of the emergent LQSS (see \cite{suppl} for details), confirms Eq. \eqref{inhroot} to describe also the NESS, provided the infinite distance limit is considered. More specifically, taking firstly $t\to+\infty$ and only after $x\to\pm\infty$, local observables are described by a free GGE based on the root density \eqref{inhroot} for $\zeta\to 0^\pm$.
It must be stressed that $\delta\rho(k)$ describes the corrections to the initial $\rho(k)$ caused by the defect, but the quasiparticle density flowing out of the impurity is rather $\rho(k)+\delta\rho(k)$. 
In order to point out the poor thermalization properties of the defect, we are going to show that $\delta \rho(k)$ can be made arbitrarily small, making the outgoing qusiparticles distribution close to $\rho(k)$, which can be chosen to be far from thermal.
In truly interacting integrable models, Eq. \eqref{inhroot} needs to be modified \cite{GHD1,GHD2}, but it retains the same physical meaning.
It is convenient to proceed through the equation of motion in the Heisenberg picture
\be\label{H_eq}
\partial_t^2\hat{\phi}_t(x)=\partial_x^2\hat{\phi}(x)-m^2\hat{\phi}(x)-:V'(\hat{\phi}(x)):\, ,
\ee
which can be equivalently reformulated in an integral equation
\be\label{int_eq}
\hat{\phi}_t(x)=\hat{\psi}_t(x)-\int_0^\infty\dd \tau\int_{-\Delta}^{\Delta} \dd y\, G_{t-\tau}(x-y) :V'(\hat{\phi}_{\tau}(y)):\, .
\ee
Above, $\hat{\psi}_t(x)$ is the field operator evolved in absence of interaction $\hat{\psi}_t(x)=e^{it\hat{H}_\text{I}}\hat{\phi}_{0}(x)e^{-it\hat{H}_\text{I}}$, $V'$ is the derivative of $V$ and $G_\tau$ the free retarded Green Function
\be\label{green}
G_\tau(x)=\Theta(t)\int_{-\infty}^\infty \frac{\dd k}{2\pi} \frac{e^{ikx}}{E(k)}\sin(E(k)t)\, .
\ee
Normal ordering ``$:\,\,:$" must be introduced to remove UV singularities \cite{peskin} and it can be achieved inserting proper counterterms in the potential, or equivalently dropping the vacuum contribution in the normal ordered correlators. In this respect, $\langle:\hat{\psi}_t(x)\hat{\psi}_{\tau}(y):\rangle$ is defined as per Eq. \eqref{corr} dropping the ``$\cos(kx)e^{-iE(k)t}/(2E(k))$" term. 

In the physical assumption that the defect's region locally relaxes to a stationary state on a finite timescale,
 the emergent LQSS and NESS can be derived from Eq. \eqref{int_eq} and $\delta \rho(k)$ \eqref{inhroot} completely determined in terms of correlation functions in the defect region. 
The time needed to the defect in order to relax contributes only as a transient, thus ineffective in the infinite time limit.
The lengthy, albeit simple, derivation is left to SM \cite{suppl} and we define
\be\label{Adef}
A_{x,x'}(t)=\lim_{T\to\infty} \langle :V'(\hat{\phi}_{t+T}(x)):\, :V'(\hat{\phi}_{T}(x')):\rangle\, ,
\ee
where the fields are computed within the defect support and the expectation values are taken with respect to the initial conditions. 
A second auxiliary function $F_{x,x'}(t)$ naturally emerges in the derivation of the LQSS \cite{suppl} and it is defined through the following convolution
\begin{multline}\label{Fdef}
\lim_{T\to \infty}\langle :V'(\hat{\phi}_{t+T}(y)):\hat{\psi}_{T}(x')\rangle=\\
\int_{-\infty}^\infty \dd \tau\, \int_{-\Delta}^{\Delta}\dd y'\, \Gamma_{-\tau}(x'-y')F_{y,y'}(t-\tau)\,.
\end{multline}
Above, $\hat{\phi}$ is always supported on the defect, while no restriction is imposed on $\hat{\psi}$. 
With these definitions, the large times emergence of the LQSS can be derived and the scattered root density $\delta \rho(k)$ computed as \cite{suppl}
\be\label{deltarho}
\delta\rho(k)=\frac{\Re(\mathcal{A}_k)-(\rho(k)+1)\Im (\mathcal{F}^+_k)+\rho(k)\Im(\mathcal{F}^-_k)}{2E(k)|v(k)|}\, ,
\ee
where
\be\label{four_A}
\mathcal{A}_k=\int_{-\infty}^\infty \dd \tau \int_{-\Delta}^{\Delta}\dd y\dd y'\,\cos\left(k(y- y')-\tau E(k)\right)A_{y,y'}(\tau)
\ee
\be\label{four_F}
\mathcal{F}_k^{\pm}=\int_{-\infty}^\infty \dd \tau \int_{-\Delta}^{\Delta}\dd y\dd y'\, e^{\pm i\left(k(y-y')-\tau E(k)\right)}F_{y,y'}(\tau)\, .
\ee

We are left with the issue of computing $A_{x,x'}(t)$ and $F_{x,x'}(t)$, which we now consider within perturbation theory.
From now on, we focus on the simplest example of a $\delta-$like defect (i.e. in Eq. \eqref{int_eq} replace $V'\to(2\Delta)^{-1} V'$ and take $\Delta\to 0$), but see SM \cite{suppl} for the finite interval case.

\section{The gaussian defect} Any perturbative analysis is constructed starting from an exact solution. Therefore, we consider a gaussian repulsive $\delta-$supported defect $V(\hat{\phi})=\mu^2\hat{\phi}^2/2$, which ultimately lays the foundation of the forthcoming perturbation theory in the truly interacting case. In Eq. \eqref{int_eq} we compute all the fields on the defect and obtain
\be\label{eq_delta}
\hat{\phi}_t(0)=\hat{\psi}_t(0)- \mu^2\int_0^\infty \dd\tau\,G_{t-\tau}(0) \hat{\phi}_\tau(0)\, .
\ee
Since we are ultimately interested in the infinite time limit and assume relaxation on the top of the defect, we can extend the time-integration domain in the infinite past.
Eq. \eqref{eq_delta} is then reformulated in the Fourier space through the definition of $g(\omega)=\int_{-\infty}^\infty \dd \tau e^{-i\omega \tau}G_\tau(0)$. From Eq. \eqref{green} we get
\be\label{g_def}
g(\omega)=\begin{cases}\text{sign}(\omega)/(2i\sqrt{\omega^2-m^2})\hspace{1pc}|\omega|> m \\
1/(2\sqrt{m^2-\omega^2}) \hspace{3pc}\,\,\,\,|\omega|<m
\end{cases}
\ee
Eq. \eqref{eq_delta} in the Fourier space states 
\be\label{eq_gauss}
\hat{\Phi}(\omega)=\hat{\Psi}(\omega)- \mu^2g(\omega)\hat{\Phi}(\omega)\, ,
\ee
where $\hat{\Phi},\hat{\Psi}$ are the Fourier transforms of the fields.
In its simplicity, Eq. \eqref{eq_gauss} has a lot to teach: even though it can be easily solved
\be\label{gauss_sol}
\hat{\Phi}(\omega)=(1+\mu^2g(\omega))^{-1}\hat{\Psi}(\omega)\, ,
\ee
it is worth to blindly proceed  through a recursive solution, as we would have done considering $\mu^2$ in perturbation theory 
\be\label{per_exp}
\hat{\Phi}(\omega)=\hat{\Psi}(\omega)- \mu^2g(\omega)\hat{\Psi}(\omega)+[\mu^2g(\omega)]^2\hat{\Psi}(\omega)+...
\ee

This series is ill defined, since $g(\omega)$ is singular when $\omega=|m|$ (despite Eq. \eqref{gauss_sol} being regular): this encloses a clear physical meaning. Tracking back the singularity from Eq. \eqref{g_def} to the definition of the Green function \eqref{green}, it is evident that the singularities are due to the modes with $k=0$, which are such that $E(k=0)=m$ and $v(k=0)=0$. 
Singularities in the frequency space are translated into secular terms when read in time: as we previously commented, secular terms are due to quasiparticles that keep on interacting as time goes further. In the defect's case, the only quasiparticles that can interact for arbitrary long times are those sat on the defect, i.e. having zero velocity, thus explaining the singularities in Eq. \eqref{per_exp}. The unperturbed field $\hat{\Psi}$ satisfies the Wick theorem and the two point function is computed on a GGE as per Eq. \eqref{corr}: a straightforward use of Eq. \eqref{gauss_sol} allows to compute $\mathcal{A}_k$ and $\mathcal{F}_k^\pm$ and subsequently $\delta\rho(k)$ 
\be\label{rho_nonint}
\delta\rho(k)=\mu^4\frac{\rho(-k)-\rho(k)}{4E^2(k)v^2(k)+\mu^4}\, .
\ee
The general structure of Eq. \eqref{rho_nonint} could have been forecast on general considerations. Being the model non interacting, $\delta\rho(k)$ must be a linear function of the initial root density $\rho(k)$. Moreover, a free particle scattering on an external potential can be either transmitted or reflected, forcing $\delta \rho(k)$ to be a function only of $\rho(\pm k)$. Finally, the parity invariance of the dynamics and the conservation of the energy flux forces $\delta\rho(k)=S(k) (\rho(k)-\rho(-k))$ with $S(k)$ symmetric in $k\to-k$, which must be determined by actual computations, as we did.

\begin{figure}[t!]
\begin{center}
\includegraphics[width=0.465\textwidth,valign=l]{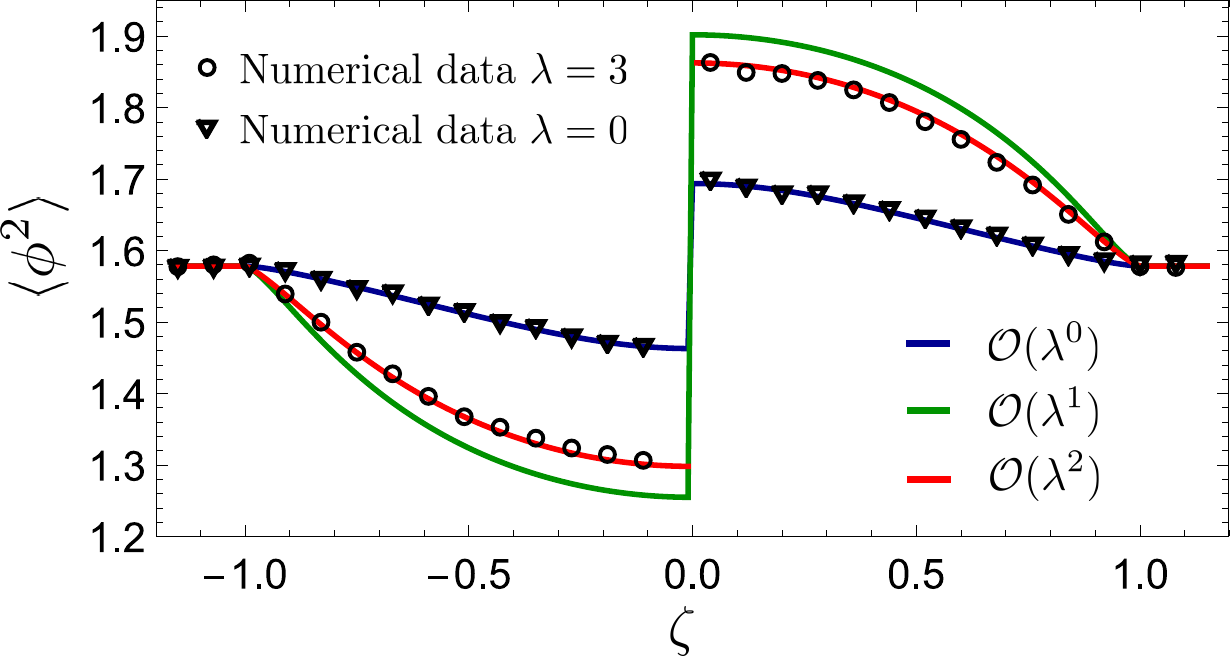}
\end{center}
\caption{\label{fig_LQSS}\emph{The analytic prediction for the LQSS is tested against the numeric simulation in the classical case, for a $\delta-$supported defect $V(\hat{\phi})=\mu^2\hat{\phi}^2/2+\lambda\hat{\phi}^4/4!$ for $\mu=1$ and $\lambda=0$ ($\mathcal{O}(\lambda^0)$ coincides with the exact result \eqref{rho_nonint}) and $\lambda=3$ (see SM \cite{suppl} for the $\mathcal{O}(\lambda^2)$ correction).
The profile of $\langle \phi^2\rangle$ as function of the ray $\zeta=x/t$ is plotted.
The initial GGE is chosen $\rho(k)=1/\big[(\beta E(k)+\beta_2 k)(e^{r(|k|-c)}+1)\big]$ with $\beta=0.5$, $\beta_2=0.4$, $r=2$, $c=20$ and bulk mass $m=1$. 
The state is a boosted thermal state, with an additional UV cut off to improve the numerical discretization \cite{suppl}. Asymmetry in $k\to-k$ in needed to have a non trivial LQSS  in the non interacting case \eqref{rho_nonint}, as well as a non trivial $\mathcal{O}(\lambda)$ correction for the interacting defect.  For $\lambda=3$, the $\mathcal{O}(\lambda^2)$ result describes the LQSS with excellent accuracy.
The continuum is discretized into $N=2^{12}$ lattice sites, with lattice spacing $\ell=0.08$ (see \cite{suppl} for the time-evolving algorithm). Roughly $30000$ realizations are used to average on the initial ensemble, a time average of the LQSS in the time interval $t\in(110,140)$ further damps statistical fluctuations. See Fig. \ref{fig_LQSS_ness_evo} for single-time screenshots.
}}
\end{figure}

\begin{figure*}[t]
\begin{center}
\includegraphics[width=0.9\textwidth,valign=l]{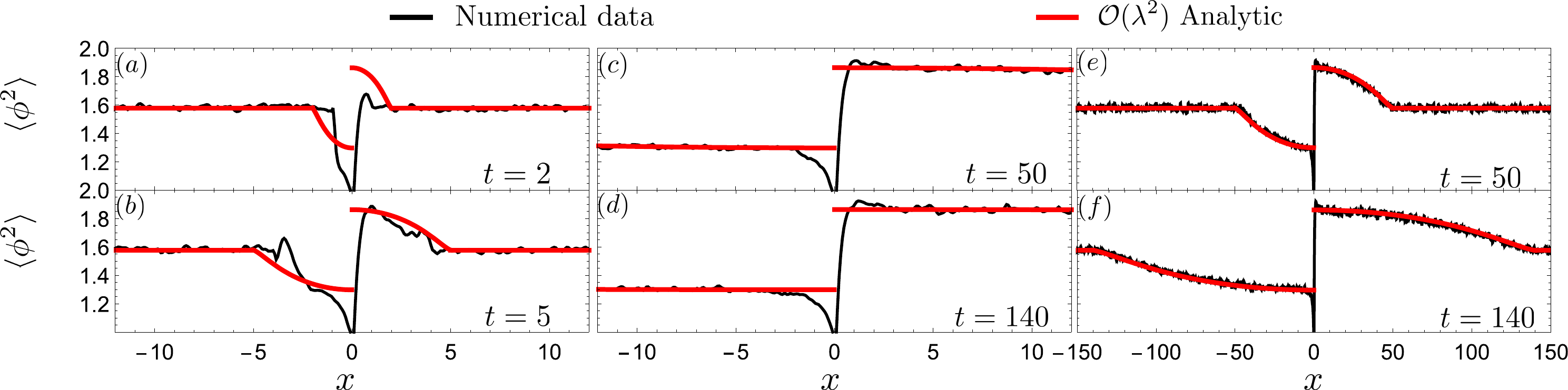}
\end{center}
\caption{\label{fig_LQSS_ness_evo}
\emph{Single-time realizations of the $\langle\phi^2\rangle$ profile for $\lambda=3$ (same parameters of Fig. \ref{fig_LQSS} are used). Panels $(a-d)$: the evolving profile is compared with the $\mathcal{O}(\lambda^2)$ analytical result (which already saturates the LQSS prediction, as depicted in Fig. \ref{fig_LQSS}) for various times and at relatively small distances from the defect, emphasizing the NESS limit. For large times, local observables attain the NESS value predicted by Eq. \eqref{inhroot}: corrections at finite distances from the impurity are observed, as expected. Panels $(e,f)$: larger distances and times are considered.
}
}
\end{figure*}
\section{The interacting defect}
We now turn on the interaction choosing $V(\hat{\phi})=\mu^2\hat{\phi}^2/2+\lambda\delta V(\hat{\phi})$, $\delta V$ being a truly interacting potential.
An expansion around $\mu^2=\lambda=0$ is plagued with singularities, exactly as it happens in Eq. \eqref{per_exp}.
However,  if we rather expand around the solution $\mu^2> 0$, $\lambda=0$, we are adding a repulsive potential on the defect, with the consequence that no quasiparticles with zero velocity can remain on the top of it. The perturbative expansion around $\mu^2>0$, $\lambda=0$ is no longer singular.
For definiteness, we focus on the explicit case $\delta V(\hat{\phi})=\hat{\phi}^4/4!$, where the analogue of Eq. \eqref{eq_gauss} is readily recast as
\begin{multline}\label{non_sing_exp}
\hat{\Phi}(\omega)=
\frac{1}{1+\mu^2 g(\omega)}\hat{\Psi}(\omega)+\\
-\frac{\lambda}{3!}\frac{g(\omega)}{1+\mu^2 g(\omega)}\int \frac{\dd^3\nu}{(2\pi)^2} \, \delta\Big(\omega-\sum_{i=1}^3\nu_i\Big):\prod_{i=1}^3\hat{\Phi}(\nu_i)\,: \,\, .
\end{multline}

A recursive solution of the above provides a $\lambda-$expansion around the solution $\mu^2>0$, $\lambda=0$ \eqref{gauss_sol}.
When compared with the perturbative expansion around $\mu^2=0$ \eqref{per_exp}, a recursive solution of Eq. \eqref{non_sing_exp} replaces the propagator $g(\omega)$ with $g(\omega)/(1+\mu^2g(\omega))$ which is no longer singular for $\mu^2>0$.
Therefore, the perturbative series remains finite at any order and secular terms are absent (see SM \cite{suppl}).
A systematic treatment of the perturbative expansion is left to SM \cite{suppl}: here we simply quote that the $\mathcal{O}(\lambda)$ result can be obtained replacing $\mu^2\to \mu^2+\lambda\alpha$ in Eq. \eqref{rho_nonint}, where
\be\label{alphadef}
\alpha=\int_0^\infty \frac{\dd k}{2\pi}  \frac{2 E(k)v^2(k)}{4 v^2(k)+\mu^4} (\rho(k)+\rho(-k))\, .
\ee
Our choice of considering a continuum model can be finally motivated: lattice systems have a bounded dispersion law, which leads to self-trapping and to the formation of boundstates even in the case of repulsive potentials \cite{self_trap}. 
This necessarily causes singularities in $g(\omega)/(1+\mu^2g(\omega))$ (associated with the bound states) that eventually plague the recursive solution of Eq. \eqref{non_sing_exp} (see SM \cite{suppl}).

\section{The classical case} So far, we have focused on a quantum model, but the Hamiltonian \eqref{HI} can be also regarded as a classical object, functional of the classical fields $\phi$ and $\Pi$.
Interestingly, in the classical realm the perturbative series can be shown to be convergent for bounded interactions \cite{suppl}.
Hereafter, we enlist the minor changes between the quantum and classical case.
The GGE correlator \eqref{corr} retains the same form even in the classical case, provided we drop the ``$\cos(kx)e^{-iE(k)t}/(2E(k))$" contribution, which in the quantum case was due to the non trivial commutator of the modes. The Green function \eqref{green} and the equation of motion \eqref{int_eq} do not change, but the normal ordering is now absent. The definitions of the $A_{x,x'}(t)$ and $F_{x,x'}(t)$ functions Eq. \eqref{Adef} and Eq. \eqref{Fdef} remain the same (without normal ordering), while in the definition of $\delta\rho(k)$ Eq. \eqref{deltarho} we must replace $\rho(k)+1\to \rho(k)$.
Incidentally, the classical and quantum results for the gaussian defect \eqref{rho_nonint} and the $\mathcal{O}(\lambda)$ order in the interacting case \eqref{alphadef} coincide.
The LQSS prediction is tested against direct numerical simulations in Fig. \ref{fig_LQSS}, finding excellent agreement, while in Fig. \ref{fig_LQSS_ness_evo} the main focus is the NESS limit.\ \\
\section{Conclusions} We considered the issue of suddenly activating an integrability-breaking localized perturbation, in an otherwise homogeneous integrable model. In contrast with the homogeneous case, which is intrinsically non-perturbative and eventually leads to thermalization, the localized impurity has less dramatic mixing properties being, at least in the example analyzed, relegated to perturbative effects. In the case where the bulk theory is free, our claim is supported by an order-by-order finite perturbative expansion constructed on top of a gaussian repulsive defect. 
Several interesting questions are left to future investigations.
First of all, non-perturbative effects are present in lattice systems, due to the phenomenon of self trapping: however, in view of our heuristic considerations, small integrability-breaking defects are expected to do not lead to thermalization, as numerically observed in \cite{fag_non_th_def}. 
Another interesting point concerns the defect's size, whose growth could lead to a crossover in its thermalizing properties.
We can expect that as the defect support is increased, the behavior of the integrability-breaking region becomes much closer to a thermodynamic system, which should thermalize due to integrability breaking.\ \\
Another interesting question concerns finite size effects: rather than a truly infinite system, we could have considered, for example, periodic boundary conditions on a ring of length $L$. In this case, a single quasiparticle can interact several times with the impurity as it travels along the ring: in this case, following our heuristic argument, non-perturbative effects are expected to build up in the infinite time limit, leading to a final thermalization. However, in the limit of large $L$, we can confide in a hydrodynamic approach generalizing Eq. \eqref{inhroot} for finite sizes and then study the relaxation to a thermal ensemble.
While we save such an analysis for future studies, we can readily comment on the expected thermalization time scale $\tau_\text{th}$. Indeed, we expect it to be proportional to the number of times a quasiparticle undergoes a scattering event, i.e. $\propto L^{-1}$, weighted with the first perturbative order of $\delta \rho$ \eqref{deltarho} which allows for non-trivial mixing among modes. As we commented, in the example we analyzed the first perturbative order simply renormalizes the mass \eqref{alphadef} of the non-interacting result, which allows only to mix modes with opposite momenta \eqref{rho_nonint}. Instead, the next-to-leading order allows for non trivial mixing, therefore we expect it can induce thermalization, leading to the estimation $\tau_\text{th}\propto 1/(\lambda^2 L)$. 
In general, quantitative results in truly interacting integrable models are surely a compelling quest.

\section{Acknowledgments} I am grateful to Maurizio Fagotti, Andrea De Luca, Bruno Bertini, Spyros Sotiriadis, Tomaz Prosen and Pasquale Calabrese for interesting discussions. I am especially indebted to Lorenzo Piroli, Bruno Bertini and Andrea De Luca for a careful reading of the manuscript and interesting comments.


\onecolumngrid
\newpage 

\setcounter{equation}{0}            
\setcounter{section}{0}             
\renewcommand\thesection{\Alph{section}}    
\renewcommand\thesubsection{\arabic{subsection}}    
\renewcommand{\thetable}{S\arabic{table}}
\renewcommand{\theequation}{S\arabic{equation}}
\renewcommand{\thefigure}{S\arabic{figure}}

\begin{center}
{\Large Supplementary Material\\ 
Lack of thermalization for integrability-breaking impurities
}
\ \\ \ \\
Alvise Bastianello
\end{center}
\ \\ \ \\ \ \\

Here we report the technical details of our analysis, organized as it follows
\begin{itemize}
\item Section \ref{sec_Fey}: representation of Eq. (6) and correlators in terms of Feynman diagrams.
\item Section \ref{sec_lqss}: large times dynamics and emergence of the LQSS, i.e. derivation of Eq. (10).
\item Section \ref{sec_delta}: $\delta-$like defect, convergence of the perturbative expansion in the classical case for certain potentials and order-by-order finitness in the quantum case.
First perturbative orders in the interacting case (i.e. those plotted in Fig. 1).
\item Section \ref{sec_ext_def}: outline of the necessary modifications needed in the case of an extended defect.
\item Section \ref{sec_gal}: galilean bosons/fermions with an interacting defect.
\item Section \ref{sec_lat}: a glimpse in lattice systems and the problem of self-trapping.
\item Section \ref{sec_num}: description of the numerical methods used to simulate the classical continuum model.
\end{itemize}

\section{Feynman diagrams}
\label{sec_Fey}

Feynman diagrams are a central tool in handling interacting systems: they constitute a remarkably compact way of representing a complicated perturbative expansion. Furthermore, partial resummations of the perturbative expansion are most easily carried out playing with the graphical representation, which sometimes gives access to non-perturbative information.
A detailed step-by-step discussion of the Feynman diagrams lays outside of the purposes of this short section, therefore we confidently assume the reader to be already familiar with the method (a complete discussion can be found in \ocite{Speskin}, as well as in several other textbooks) and outline the Feynman rules we need.

Our ultimate goal is a Feynman-diagram representation of the field correlation functions, which can be achieved in two subsequent steps
\begin{enumerate}
\item Represent through Feynman diagrams the iterative solution of the integral equation Eq. (6), reported hereafter for convenience
\be\label{S_inteq}
\hat{\phi}_t(x)=\hat{\psi}_t(x)-\int_0^\infty\dd \tau\int_{-\Delta}^{\Delta} \dd y\, G_{t-\tau}(x-y) :V'(\hat{\phi}_{\tau}(y)):\, .
\ee
\item Compute the correlation functions taking as an input the previous step.
\end{enumerate}

For the time being we work in real space and time and with no restrictions on the coordinate domain: trivial modifications allow to remove the lower bound to time integration and consider $\tau\in (-\infty,\infty)$ rather than $\tau\in(0,\infty)$ and switch to the Fourier space, if needed.
Assume for simplicity $V(\hat{\phi})=\frac{\lambda}{n!}\hat{\phi}^n$, generalizations to arbitrary Taylor expandable potentials will appear clear. Since $V'(\hat{\phi})=\frac{\lambda}{(n-1)!}\hat{\phi}^{n-1}$, we represent each interaction by mean of a vertex with $n$ departing legs. The Green function $G$ is associated with a dashed line.
The recursive solution of Eq. \eqref{S_inteq} is then represented through all the possible tree-like diagrams (i.e. no loops) constructed with the following rules
\begin{itemize}
\item External legs are such that one (dashed) is associated with the desired solution $\hat{\phi}_t(x)$, the others to the unperturbed solutions $\hat{\psi}$ computed at the time and position of the vertex to which they are attached.
\item Internal legs are mediated by the dashed lines associated with the Green function that are thus attached to two vertexes. The coordinates appearing in $G$ are those of the two vertexes.
\item The $1/(n-1)!$ contribution of each vertex is (almost) canceled by the sum of several equivalent diagrams. In this perspective, each interaction vertex contributes simply as $-\lambda$.
\item Each graph must be divided by an overall symmetry factor, which is equal to the number of permutations of legs which do not change the topology of the graph.
\item After an integration of the space/time coordinates of the vertexes on the suitable domain, $\hat{\phi}_t(x)$ is obtained summing over all the possible graphs.
\end{itemize}

\begin{figure}[t]
\begin{center}
\includegraphics[width=1\textwidth]{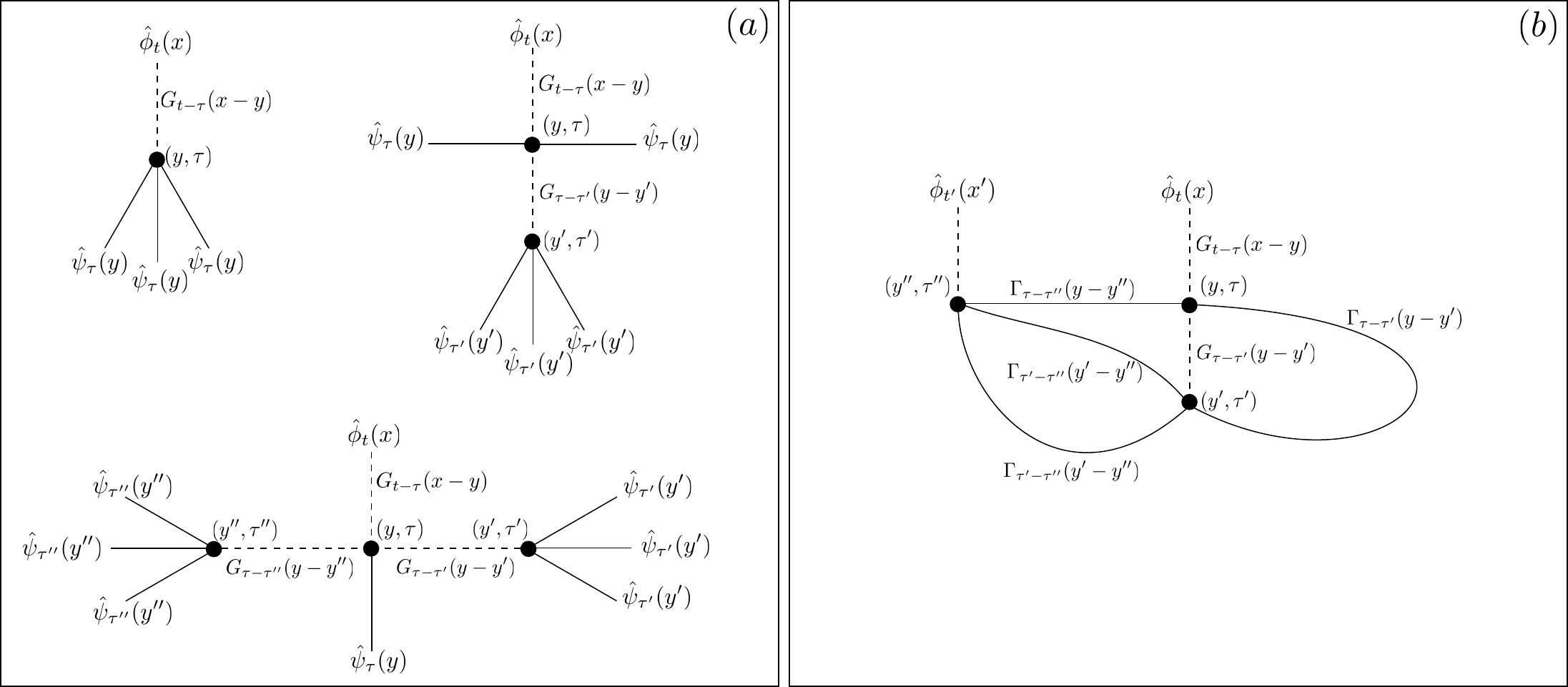}
\end{center}
\caption{\label{fig_feyn1}\emph{$(a)$ Tree level Feynman diagrams appearing in the recursive solution of Eq. \eqref{S_inteq} for $V(\hat{\phi})=\lambda\hat{\phi}^4/4!$, $(b)$ Feynman diagrams representing correlators: these are obtained joining together the external legs of the graphs representing the iterative solution of Eq. \eqref{S_inteq}. As an example, we construct a graph joining together the first two diagrams of panel $(a)$.
}}
\end{figure}

An example is depicted in Fig. \ref{fig_feyn1}$(a)$.
From these Feynman diagrams we can now construct those of the correlation functions: consider for example the case of the two point correlator $\langle \hat{\phi}_t(x)\hat{\phi}_{t'}(x')\rangle$, the generalization to multipoint correlators will appear trivial.
Within a recursive solution of Eq. \eqref{S_inteq}, correlators of $\hat{\phi}_t(x)$ are obtained by mean of a repetitive use of the Wick theorem on the fields $\hat{\psi}$. Therefore, the Feynman diagrams associated with the correlator can be constructed as it follows: choose one Feynman diagram in the representation of $\hat{\phi}_t(x)$ and one concerning $\hat{\phi}_{t'}(x')$, then connect pairwise together all the possible external lines associated with the fields $\hat{\psi}$.
These new internal lines are associated with the correlator $\langle \hat{\psi}\hat{\psi}\rangle$, therefore with $\Gamma $ (3). $\Gamma$ must be evaluated at positions an times equal to the difference of the coordinates of the two vertexes which are connected by $\Gamma$. Normal ordering on $\Gamma$ must be used if the line starts and ends at the same vertex.
An example is depicted in Fig. \ref{fig_feyn1}$(b)$.
Finally, all the possible choices among the Feynman diagrams contributing to $\hat{\phi}$ must be considered and connected together in all the possible ways: this generates all the Feynman diagrams associated with the correlator.

In this respect, it is convenient to reconsider the symmetry factor as it follows: compute the contribution of the Feynman diagram of the single field $\hat{\phi}$ ignoring the symmetry factors and, after the Feynman diagram of the correlator has been constructed, divide by the number of permutations of leg and vertexes which leave the diagram the same. Finally, the expansion of the correlator is obtained summing over all the distinct Feynman graphs.

\section{The Local Quasi-Stationary State}
\label{sec_lqss}

The goal of this section is to show the emergence of the LQSS and express the root density leaving the impurity in terms of the correlation functions on top of the defect, i.e. Eq. (10) together with the definitions Eq. (11-12).
This relation is unperturbative and relays on the assumption that the correlators on the impurity reach a steady state at large times: in this case the large times behavior (i.e. the LQSS) will be completely determined in terms of the mentioned correlators. 

Hereafter, we consider the two point correlator $\langle \hat{\phi}_t(x)\hat{\phi}_{t'}(x')\rangle$ at large times and far from the defect, recognizing that it can be written as if the ensemble was homogeneous, but with a space-time dependent root density. Furthermore, such an inhomogeneous root density will be in the form Eq. (4).
For simplicity we consider the equaltime correlator $t=t'$, but the same analysis can be performed on $t\ne t'$, leading to the same conclusions.
In principle, the gaussification of the multipoint correlators (i.e. the validity of the Wick theorem) must be checked: this can be done, but it requires further lengthy calculations closely related to those presented in Ref. \ocite{Sgaussification1}.

Using the exact integral equation \eqref{S_inteq} we can surely write
\begin{multline}\label{eq:2pcor}
\langle\hat{\phi}_t(x)\hat{\phi}_{t}(x')\rangle=\Bigg\langle\bigg(\hat{\psi}_t(x)-\int_0^\infty \dd \tau\, \int_{-\Delta}^{\Delta}\dd y\, G_{t-\tau}(x-y):V'(\phi_\tau(y)):\bigg) \\ 
\bigg(\hat{\psi}_{t}(x')-\int_0^\infty \dd\tau\, \int_{-\Delta}^{\Delta} \dd y'\, G_{t-\tau'}(x'-y'):V'(\phi_{\tau'}(y')):\bigg)\Bigg\rangle
\end{multline}

We expand the above and consider each term separately
\begin{multline}\label{eq:twoexp}
\langle\hat{\phi}_t(x)\hat{\phi}_{t}(x')\rangle=\langle\hat{\psi}_t(x)\hat{\psi}_{t}(x')\rangle+\\
-\int_0^\infty \dd\tau'\, \int_{-\Delta}^{\Delta}\dd y'\, G_{t-\tau'}(x'-y')\langle \hat{\psi}_t(x):V'(\hat{\phi}_{\tau'}(y')):\rangle-\int_0^\infty \dd\tau\, \int_{-\Delta}^{\Delta}\dd y\, G_{t-\tau}(x-y)\langle :V'(\hat{\phi}_{\tau}(y)):\hat{\psi}_{t}(x')\rangle+ \\
+\int_0^\infty \dd\tau \dd\tau'\, \int_{-\Delta}^{\Delta}\dd y\dd y'\, G_{t-\tau}(x-y)G_{t-\tau'}(x'-y')\langle :V'(\hat{\phi}_\tau(y))::V'(\hat{\phi}_{\tau'}(y')):\rangle
\end{multline}

Consider now the last row: we assume that after a finite time the correlator $\langle :V'(\hat{\phi}_\tau(y))::V'(\hat{\phi}_{\tau'}(y')):\rangle$ has reached a steady state and define $A_{x,x'}(t)$ as per Eq. (8), which we ultimately replace in the above. Any integration over a finite time window contributes as a transient with respect to the infinite time limit.

As a further technical assumption, we require the correlators on the defect to decorrelate when separated by an infinite time $\lim_{|\tau|\to\infty}A_{x,x'}(\tau)=0$ and we assume $A_{x,x'}(\tau)$ approaches zero fast enough in order to have an integrable Fourier transform. 

Subsequently, we consider a large distance-time expansions: in this respect, it is useful to consider the asymptotics of the Green function extracted by mean of a saddle point approximation
\be\label{saddle_G}
G_t(x)=\Theta(t)\int_{-\infty}^\infty \frac{\dd k}{2\pi}\frac{\sin(t E(k) -kx)}{E(k)}\simeq\frac{\Theta(t)}{2\pi}\frac{1}{E(k_\zeta)}\sqrt{\frac{2\pi}{t\partial_k v(k_\zeta)}}\sin\Big(tE(k_\zeta) -k_\zeta x+\pi/4\Big)+\mathcal{O}(t^{-1})
\ee
Above, $k_\zeta$ is the solution of the equation $\zeta=x/t=v(k_\zeta)$ where we recall $v(k)=\partial_k E(k)$ is the group velocity. 
Using Eq. \eqref{saddle_G} and with some tedious, but straightforward, calculations we find
\ba\label{eq:LQSSA}
&&\int_0^\infty \dd\tau \dd\tau'\, \int_{-\Delta/2}^{\Delta/2}\dd y\dd y'\, G_{t-\tau}(x-y)G_{t-\tau'}(x'-y')\langle :V'(\hat{\phi}_\tau(y))::V'(\hat{\phi}_{\tau'}(y')):\rangle\simeq\\
&&\nonumber\int_{-\infty}^\infty\frac{\dd k}{2\pi}\frac{\Theta(\zeta v(k))\Theta(|v(k)|-|\zeta|)\delta\rho_A(k)}{E(k)}\cos\left(k(x-x')\right)
\ea
where
\be\label{rho_A}
\delta\rho_A(k)=\frac{\Re (\mathcal{A}_k)}{2E(k)|v(k)|}
\ee
with $\mathcal{A}_k$ given in Eq. (11).

\begin{figure}[t]
\begin{center}
\includegraphics[width=0.9\textwidth]{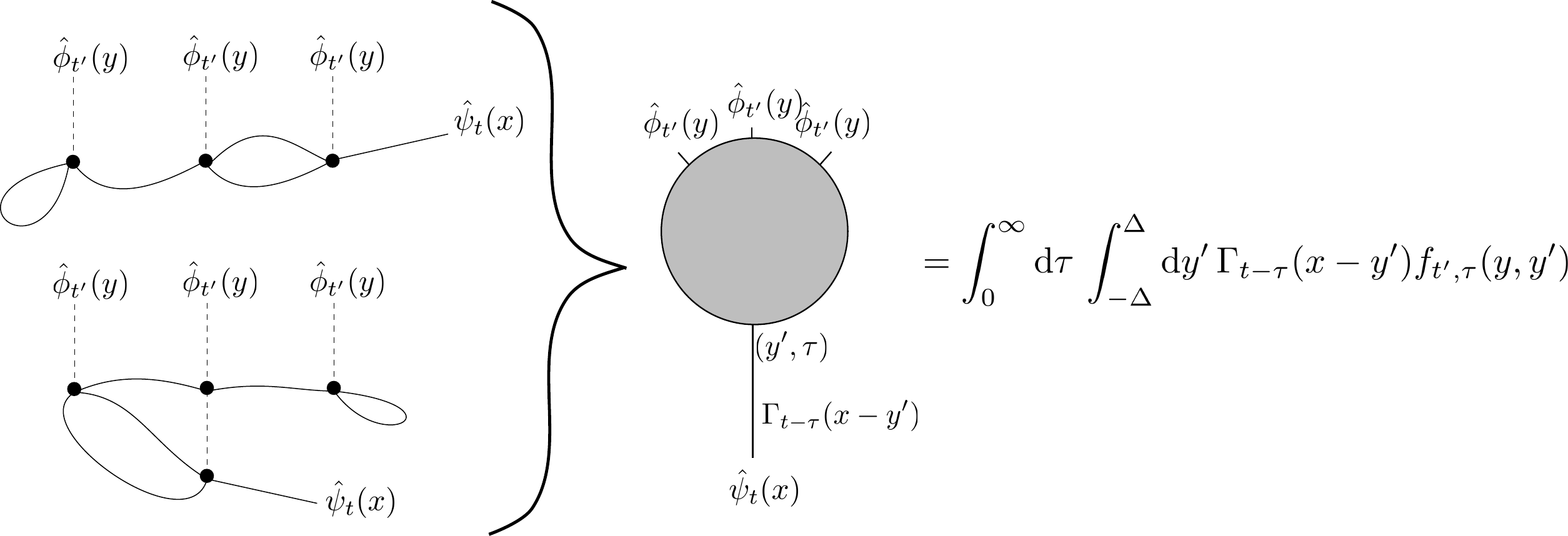}
\end{center}
\caption{\label{diag_F}\emph{Feynman diagrams representing $\langle : V'(\hat{\phi}_{t'}(y)):\hat{\psi}_t(x)\rangle$, where we used as an example $V'(\hat{\phi})=\lambda\hat{\phi}^3/3!$.
All the possible diagrams we can draw have the same structure, which is then promoted to be an exact identity.
}}
\end{figure}

We can now consider the remaining terms in Eq. \eqref{eq:twoexp}, i.e. the second row. Here, provided we assume the validity of Eq. (9) (which will be soon justified), we can repeat the same calculations using the large distance expansion of $\Gamma$
\begin{multline}\label{Sgamma_exp}
\Gamma_{t}(x)=\int_{-\infty}^\infty \frac{\dd k}{2\pi}\frac{\cos(E(k)t-kx)}{E(k)}\rho(k)+\frac{\cos(kx)}{2E(k)}e^{-iE(k)t}\simeq\\
 \frac{1}{2\pi}\frac{1}{2E(k_\xi)}\sqrt{\frac{2\pi}{t\partial_k v(k_\zeta)}}\left(e^{-i(tE(k_\zeta)-k_\zeta x+\pi/4)}(\rho(k_\zeta)+1)+e^{i(tE(k_\zeta)-k_\zeta x+\pi/4)}\rho(k_\zeta)\right)
\end{multline}
and find that the second row of Eq. \eqref{eq:twoexp} can be written in the same form of Eq. \eqref{eq:LQSSA}, provided we replace $\delta\rho_A\to\delta\rho_F$
\be
\delta\rho_F(k)=\frac{-(\rho(k)+1)\Im(\mathcal{F}^+_k)+\rho(k)\Im(\mathcal{F}^-_k)}{2|v(k)|E(k)}
\ee
with $\mathcal{F}_k^\pm$ defined in Eq. (12). Summing all the contributions we readily recognize the two point correlator $\langle\hat{\phi}_t(x)\hat{\phi}_{t}(x')\rangle$ to acquire the LQSS form with $\delta \rho(k)=\delta\rho_A(k)+\delta\rho_F(k)$
\be\label{Sdeltarho} 
\delta\rho(k)=\frac{\Re(\mathcal{A}_k)-(\rho(k)+1)\Im (\mathcal{F}^+_k)+\rho(k)\Im(\mathcal{F}^-_k)}{2E(k)|v(k)|}\, ,
\ee
which is Eq. (10) reported in the main text.

The validity of Eq. (9) can be justified at any order in the diagrammatic expansion as it follows: consider the diagrams for $\langle : V'(\hat{\phi}_{t'}(y)):\hat{\psi}_t(x)\rangle$, which are obtained expanding $:V'(\hat{\phi}_{t'}(y)):$ and then contracting the resulting graphs with $\hat{\psi}_t(x)$ (see Fig. \ref{diag_F}). When the field $\hat{\psi}_t(x)$ is contracted with a $\hat{\psi}$ field contained in the expansion of $:V'(\hat{\phi}_{t'}(y)):$, the latter is always constrained on the defect support. Therefore, any Feynman diagram in the expansion of $\langle : V'(\hat{\phi}_{t'}(y)):\hat{\psi}_t(x)\rangle$ can be written in the following form, that we promote to be an identity of the correlator itself
\be
\langle :V'(\hat{\phi}_{t'}(y)):\hat{\psi}_{t}(x)\rangle=
\int_0^\infty \dd \tau\, \int_{-\Delta}^{\Delta}\dd y'\, \Gamma_{t-\tau}(x-y')f_{t',\tau}(y,y')\, ,
\ee
where the function $f_{t,\tau}(y,y')$ contains the contribution of all the Feynman diagrams. If we require the correlator to reach a stationary state on the defect we are forced to require \emph{i)} $f_{t',\tau}(y,y')$ to become time translational invariant $f_{t',\tau}(y,y')=f_{t'+T,\tau+T}(y,y')$ and \emph{ii)} decaying fast enough in $|t-\tau|\to\infty$ in such a way we can safely extend the time integration from $\tau\in[0,+\infty)$ to $\tau\in(-\infty,\infty)$. In this case, we are naturally lead to Eq. (9).

We quickly comment on the fact that the same calculations can be repeated in the classical case, with minor modifications. 
As already commented, the term ``$\cos(kx)e^{-iE(k)t}/(2E(k))$" in the first line of Eq. \eqref{Sgamma_exp} comes from the non trivial commutation relations of the quantum modes, thus it is absent in the classical realm. Subsequently, in the large distance expansion (second line of Eq. \eqref{Sgamma_exp}) we should replace $\rho(k)+1\to \rho(k)$, which ultimately implies the same substitution in the definition of $\delta \rho(k)$ Eq.\eqref{Sdeltarho}.
Furthermore, in the classical case $\mathcal{A}_k$ is real and $\mathcal{F}^+_k=\big[\mathcal{F}^-_k\big]^*$.

\section{The $\delta-$defect}
\label{sec_delta}

In this section we analyze the $\delta-$like defect, discuss the convergence of the perturbative expansion in the classical case and provide the expression for the first perturbative orders displayed in Fig. 1. As already stressed in the main text, the perturbative expansion can be convergent only if we expand around a repulsive potential placed on the defect. In this respect, we assume ($\mu^2>0,\lambda>0$)
\be
V(\hat{\phi})=\frac{\mu^2}{2}\hat{\phi}^2+\lambda \delta V(\hat{\phi})\,
\ee
where $\delta V$ contains the truly interacting part which, for the time being, is left arbitrary. The exact integral equation describing the solution to the equation of motion is therefore
\be\label{Sint_delta}
\hat{\phi}_t(0)=\hat{\psi}_t(0)- \mu^2\int_0^\infty \dd\tau\,G_{t-\tau}(0) \hat{\phi}_\tau(0)-\lambda\int_0^\infty \dd\tau\,G_{t-\tau}(0) :\delta V'(\hat{\phi}_\tau(0)):\, .
\ee

A naive recursive solution is equivalent to an expansion around the solution $\mu^2=\lambda=0$ that is not what we are looking for. Therefore, we define a new Green function $\tilde{G}_t$ satisfying
\be\label{G_def}
\tilde{G}_t=G_t(0)-\mu^2\int_{-\infty}^\infty \dd\tau\, G_{t-\tau}(0)\tilde{G}_{\tau}\, .
\ee
The function $\tilde{G}_t$ is nothing else than the Green function (computed on the defect) associated with the equation of motion in presence of the gaussian defect, i.e. $\mu^2\ne 0$ and $\lambda=0$. 

Furthermore, we define $\varphi_t$ as the solution of
\be
\hat{\varphi}_t=\hat{\psi}_t(0)- \mu^2\int_{0}^\infty \dd\tau\,G_{t-\tau}(0)\hat{\varphi}_\tau\, .
\ee

In terms of these newly introduced quantities, Eq. \eqref{Sint_delta} is rewritten as
\be\label{def_int}
\hat{\phi}_t(0)=\hat{\varphi}_t-\lambda\int_0^\infty \dd\tau\,\tilde{G}_{t-\tau} :\delta V'(\hat{\phi}_\tau(0)):\, .
\ee

\subsection{Convergence in the classical realm}
\label{conv_cl}

We can now easily discuss the convergence of the recursive solution of Eq. \eqref{def_int} in the classical case for a certain class of potentials. Through this section, the field $\hat{\phi}\to \phi$ is a simple function rather than an operator.

An upper bound to the convergence radius of the expansion can be given in the assumption of \emph{i)} bounded interaction $|\delta V'(\phi)|\le C_1$ and \emph{ii)} H\"older condition $|\delta V'(\phi)-\delta V'(\phi')|\le C_2|\phi-\phi'|$. 

We look closely at the Green function $\tilde{G}$ and introduce its Fourier transform $\tilde{g}(\omega)=\int_{-\infty}^\infty \dd \tau e^{-i\omega \tau}\tilde{G}_\tau$. The defining equation \eqref{G_def} is easily solved in the Fourier space
\be\label{tilde_g_four}
\tilde{g}(\omega)=\frac{g(\omega)}{1+\mu^2g(\omega)}=\begin{cases}\frac{1}{\text{sign}(\omega)2i\sqrt{\omega^2-m^2}+\mu^2}\hspace{1pc}&|\omega|\ge m \\ \, \\
\frac{1}{2\sqrt{m^2-\omega^2} +\mu^2}\hspace{3pc}\,\,\,\,&|\omega|< m
\end{cases}
\ee
where, in the last equality, we used the expression of $g(\omega)$ Eq. (14). The removal from $\tilde{g}(\omega)$ of the singularity that was present in $g(\omega)$ changes the large time behavior of the Green function
\be\label{G_larget}
G_t(0)\sim t^{-1/2}\hspace{3pc} \tilde{G}_t\sim t^{-3/2}\, .
\ee
In particular, the $L^1$ norm of $\tilde{G}_t$ is finite
\be
\int_{-\infty}^\infty \dd t\,  |\tilde{G}_t|=M\le \infty\, .
\ee
Consider now the recursive solution of Eq. \eqref{def_int}
\be
\phi^{(n+1)}_t(0)=\varphi_t-\lambda\int_0^\infty \dd\tau\,\tilde{G}_{t-\tau} \delta V'(\phi^{(n)}_\tau(0))\, ,
\ee
with $\phi^{(0)}_t(0)=\varphi_t$. 
From the above definition we can readily construct the following chain of inequalities
\begin{multline}\label{conv_bound}
\max|\phi_t^{(n+1)}(0)-\phi_t^{(n)}(0)|\le \lambda\max\Big|\int_0^\infty \dd\tau\,\tilde{G}_{t-\tau} \big[\delta V'(\phi^{(n)}_\tau(0))-\delta V'(\phi^{(n-1)}_\tau(0))\big]\Big|\\
\le \lambda\max\left(\int_0^\infty \dd\tau\,|\tilde{G}_{t-\tau}|\right)\max\Big| \delta V'(\phi^{(n)}_t(0))-\delta V'(\phi^{(n-1)}_t(0))\Big|\le \lambda M C_2 \max|\phi_t^{(n)}(0)-\phi_t^{(n-1)}(0)|\, .
\end{multline}

Then
\begin{multline}\label{ser_conv}
\max|\phi_t^{(n+N)}(0)-\phi_t^{(n)}(0)|=\max\Big|\sum_{j=n}^{N-1+n}\big(\phi_t^{(j+1)}(0)-\phi_t^{(j)}(0)\big)\Big|\le \sum_{j=n}^{N-1+n}\max|\phi_t^{(j+1)}(0)-\phi_t^{(j)}(0)|\\
\le\max|\phi_t^{(n+1)}(0)-\phi_t^{(n)}(0)|\sum_{j=0}^{N-1}(\lambda M C_2)^j
\end{multline}

The limit $\lim_{n\to\infty} \phi_t^{(n)}(0)$ is thus guaranteed to exist (and finite) if the following geometric series converges
\be
\sum_{j=0}^\infty (\lambda M C_2)^j\le \infty\, ,
\ee
which is true as long as $\lambda M C_2< 1$, leading to the estimated convergence radius $\lambda\le 1/(M C_2)$.

\subsection{The first perturbative orders}
\label{pert_ord}

Here we discuss the first orders in the perturbative expansion, for definitness we focus on the interaction we considered in the main text, i.e.
\be\label{deltaV_phi4}
\delta V(\hat{\phi})=\frac{1}{4!}\hat{\phi}^4\, .
\ee

Unfortunately, in the classical case such an interaction does not satisfy the conditions assumed in the previous section in order to prove the convergence of the perturbative series. However, we will immediately understand that at least any order in the $\lambda-$expansion is finite, both in the quantum and in the classical case.

Since we aim to compute the correlators on the top of the defect and in the infinite time limit, we can equivalently consider Eq. \eqref{def_int} and extend the time integration to the whole real axis, i.e.
\be\label{int_inft}
\hat{\phi}_t(0)=\hat{\varphi}_t-\lambda\int_{-\infty}^\infty \dd\tau\,\tilde{G}_{t-\tau} :\delta V'(\hat{\phi}_\tau(0)):\, .
\ee

Notice that, recasting the above in the Fourier space and choosing $\delta V$ as per Eq. \eqref{deltaV_phi4}, we readily recover Eq. (19) of the main text.
\be
\hat{\Phi}(\omega)=
\frac{1}{1+\mu^2 g(\omega)}\hat{\Psi}(\omega)
-\frac{\lambda}{3!}\frac{g(\omega)}{1+\mu^2 g(\omega)}\int \frac{\dd^3\nu}{(2\pi)^2} \, \delta\Big(\omega-\sum_{i=1}^3\nu_i\Big):\prod_{i=1}^3\hat{\Phi}(\nu_i)\,: \,\,.
\ee

In principle, we should now compute $ \langle :V'(\hat{\phi}_{t}(0)):\, :V'(\hat{\phi}_{t'}(0)):\rangle$ and $\langle :V'(\hat{\phi}_{t}(0)):\hat{\psi}_{t'}(0)\rangle$, then from these extract the functions $A$ and $F$ from their definitions Eq. (8-9). However, the needed Feynman diagrams are complicated even at the first orders.
In this respect, it is more convenient to express the desired correlators in terms of simpler correlation functions, using the integral equation \eqref{int_inft}.

For example, we readily obtain the identity
\be\label{eq_four}
\langle \hat{\phi}_t(0) \hat{\psi}_{t'}(0)\rangle=\langle \hat{\varphi}_t \hat{\psi}_{t'}(0)\rangle-\lambda\int_{-\infty}^\infty \dd\tau\,\tilde{G}_{t-\tau} \langle:\delta V'(\hat{\phi}_\tau(0)): \hat{\psi}_{t'}(0)\rangle\, .
\ee

Taking the Fourier transform and defining (notice that $\langle \hat{\phi}_t(0) \hat{\psi}_{t'}(0)\rangle=\langle \hat{\phi}_{t+T}(0) \hat{\psi}_{t'+T}(0)\rangle$, since we are considering the stationary state attained in the infinite time limit)
\be\label{d_def}
d(\omega)=\int_{-\infty}^\infty \dd t \,e^{-i\omega (t-t')} \langle \hat{\phi}_t(0) \hat{\psi}_{t'}(0)\rangle\, .
\ee
Then Eq. \eqref{eq_four} is readily recast as
\be
d(\pm E(k))=\frac{\gamma(\pm E(k))}{1+\mu^2 g(\pm E(k))}-\frac{g(\pm E(k))}{1+\mu^2 g(\pm E(k))}\left[\gamma(\pm E(k))\mathcal{F}^\pm_k-\mu^2 d(\pm E(k))\right]
\ee
where we used $V'(\hat{\phi})=\mu^2\hat{\phi}+\lambda \delta V'(\hat{\phi})$ and defined
\be
\gamma(\omega)=\int_{-\infty}^\infty \dd t \, e^{-i\omega t}\Gamma_t(0)=\begin{cases} 0&\hspace{3pc} |\omega|<m \\ \, \\ \frac{1}{2\sqrt{\omega^2-m^2}}\Big[\rho\big(\sqrt{\omega^2-m^2}\big)+\rho\big(-\sqrt{\omega^2-m^2}\big)+2\Theta(-\omega)\Big]&\hspace{3pc} |\omega|\ge m\end{cases}\, ,
\ee
where $\Gamma_t(x)$ is defined in Eq. (3) (in the classical case or if normal ordering is considered, the ``$2\Theta(-\omega)$" term is absent).
Solving for $\mathcal{F}_k$ we finally get
\be\label{F_rel}
\mathcal{F}_k^\pm=\frac{\gamma(\pm E(k))-d(\pm E(k))}{g(\pm E(k))\gamma(\pm E(k))}\, .
\ee
With similar passages and defining $c(\omega)$ as
\be\label{c_def}
c(\omega)=\int_{-\infty}^\infty \dd t \,e^{-i\omega (t-t')} \langle \hat{\phi}_t(0) \hat{\phi}_{t'}(0)\rangle
\ee
we obtain a further identity 
\be\label{A_rel}
\mathcal{A}_k=\frac{1}{2g(E(k))g(-E(k))}\Bigg[c(E(k))+c(-E(k))-\gamma(E(k))-\gamma(-E(k))+2g(E(k))\gamma(E(k))\mathcal{F}^+_k+2g(-E(k))\gamma(-E(k))\mathcal{F}^-_k \Bigg]\, .
\ee

We are thus left with the simpler problem of computing certain two point functions $c(\omega)$ and $d(\omega)$, then from these $\mathcal{A}_k$ and $\mathcal{F}_k^\pm$ easily follow.
Notice that from the definitions of $c(\omega)$ and $d(\omega)$ Eq. \eqref{d_def} and Eq. \eqref{c_def} we clearly have
\be\label{cd_rel}
\langle\hat{\Phi}(\omega)\hat{\Phi}(\omega') \rangle=2\pi \delta(\omega+\omega')c(\omega)\, ,\hspace{4pc}\langle\hat{\Phi}(\omega)\hat{\Psi}(\omega') \rangle=2\pi \delta(\omega+\omega')d(\omega)\, .
\ee

The Feynman diagrams needed to describe the solution of Eq. \eqref{eq_four} and the correlators can be constructed similarly to what we did in Section \ref{sec_Fey}, however they are best described in the Fourier space. The rules are the following.
The perturbative solution of $\hat{\Phi}(\omega)$ is represented as tree-like Feynman diagrams. The interaction vertex is a dot with $4-$ departing legs, which can be either dashed or continuum.

\begin{itemize}
\item One external leg is associated with $\hat{\Phi}(\omega)$, the others to the fields $\hat{\Psi}$. At this level, all the internal lines are dashed lines and are associated with the Green function $\tilde{G}_t$, which in the Fourier space is simply $\tilde{g}(\omega)=g(\omega)/(1+\mu^2 g(\omega))$. The external line associated with $\hat{\Phi}(\omega)$ is a dashed line, while all the others are continuum lines.
\item In order to construct the correlator $\langle \hat{\Phi}(\omega)\hat{\Phi}(\omega')\rangle$, draw together two of the diagrams described above and join pairwise all the continuum lines. When computing $\langle \hat{\Phi}(\omega)\hat{\Psi}(\omega')\rangle$, draw a Feynman diagram associated with $\hat{\Phi}(\omega)$ and connect all the continuum lines except one, which remains an external leg associated with $\hat{\Psi}(\omega')$.
\item Each line carries its own frequency $\nu$ and it must have its own direction, along which $\nu$ flows. Continuous internal lines are associated with $\tilde{\gamma}(\nu)=\gamma(\nu)/|1+\mu^2g(\nu)|^2$, while the external continuous line associated with $\hat{\Psi}(\omega')$ contributes as $\gamma(\omega')/(1+\mu^2 g(\omega'))$.
\item Each vertex carries a $-\lambda 2\pi\delta(\sum_i \pm_i \nu_i)$ factor, where $\nu_i$ are the frequencies flowing into the vertex. The sign $\pm_i$ is chosen $+$ if the frequency flows into the vertex, $-$ in the other case.
\item Integrate over all frequencies $\nu_i$ of the graph, with a measure $\dd \nu_i/(2\pi)$.
\item Divide for the symmetry factor, i.e. the number of permutations of internal legs which leave the graph unchanged, and sum all the possible diagrams.
\end{itemize}

\begin{figure}[t]
\begin{center}
\includegraphics[width=1\textwidth,valign=l]{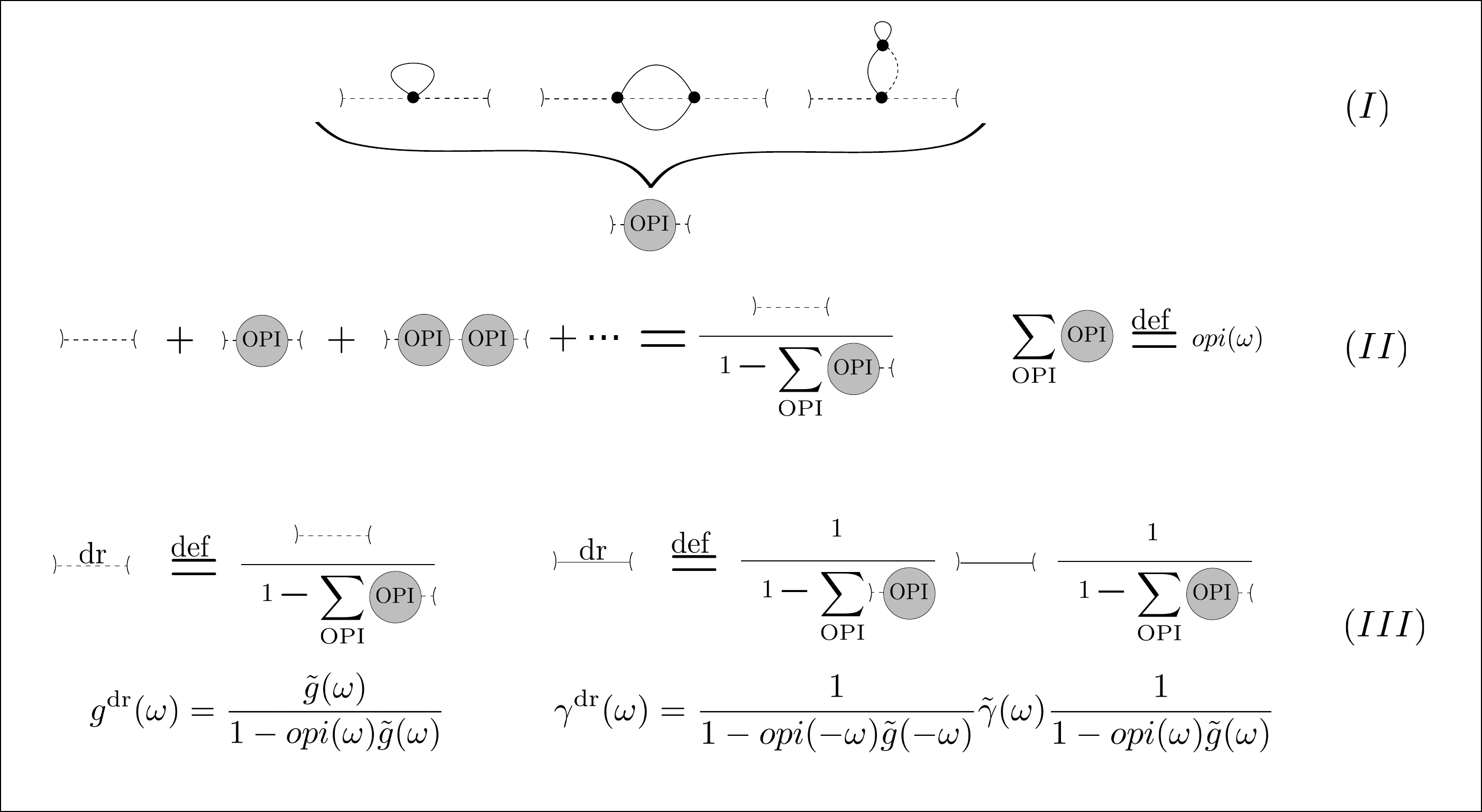}
\end{center}
\caption{\label{fig_OPI}\emph{A few examples of OPIs (I).
Several graphs can be constructed attaching together OPIs and forming ``chains" of arbitrary length (II): the contribution of these graphs can be easily resummed as a geometric series, which can be absorbed in a dressing of $\tilde{g}$ and $\tilde{\gamma}$ (III).
}}
\end{figure}

\begin{figure}[b]
\begin{center}
\includegraphics[width=0.2\textwidth,valign=l]{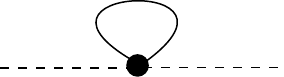}
\end{center}
\caption{\label{fig_I_ord}\emph{The only $\mathcal{O}(\lambda)$ graph that can be drawn has a single interaction vertex, a loop and two departing legs. Above we represented the departing legs as dashed lines, but one of the two can also be chosen to be continuous. This is clearly an OPI graph.
}}
\end{figure}

With these rules we can systematically compute the contribution up to the desired order in $\lambda$. We stress that each graph is constructed using as building blocks $\tilde{\gamma}(\nu)$ and $\tilde{g}(\nu)$, both of them being non singular and decaying as $\sim \nu^{-1} $ for large frequencies: once normal ordering has been considered, all the integrals are UV convergent. Thus, all the terms of the expansions are finite.

The formalism of Feynman diagrams can be even pushed further, performing partial resummations of the graphs.

In this respect, a key role is played by the one-particle irreducible (OPI) graphs. We borrow the same notation of high-energy physics and the reader can refer to Ref. \ocite{Speskin} for a pedagogical and extensive discussion of diagrammatic techniques and OPI resummation.
The OPIs in our case are defined as ``blocks" such that they can be disconnected from the whole diagram cutting two dashed lines. Moreover, these blocks cannot be further disconnected cutting a single additional dashed line: a few examples are given in Fig. \ref{fig_OPI}.
Several graphs can be obtained composing together OPIs (Fig. \ref{fig_OPI}): because of frequency conservation, the frequencies in the incoming and outgoing legs of an OPI must be the same. This enormously simplifies evaluating the blocks constructed composing together several OPIs, whose sum is a simple geometric series and can be absorbed in a ``dressing" of the dashed and continuous lines (see again Fig. \ref{fig_OPI}).
When computing Feynman diagrams, we can replace the continuum and dashed lines with their dressed counterparts: of course, in the ``dressed" Feynman diagrams we should avoid any OPI, since their contribution has already been taken in account.

We can finally use the constructed machinery to compute the needed correlators, for the time being we focus on the $\mathcal{O}(\lambda)$ contribution. At this order, we can construct only one Feynman diagram (Fig. \ref{fig_I_ord}), which is clearly an OPI: using this graph in the dressing procedure, we get 
\be
\langle \hat{\Phi}(\omega)\hat{\Phi}(\omega')\rangle=2\pi\delta(\omega+\omega')\frac{1}{1+\lambda \alpha\tilde{g}(-\omega)}\tilde{\gamma}(\omega)\frac{1}{1+\lambda \alpha\tilde{g}(\omega)}+\mathcal{O}(\lambda^2)
\ee
\be
\langle \hat{\Phi}(\omega)\hat{\Psi}(\omega')\rangle=2\pi\delta(\omega+\omega')\frac{1}{1+\lambda \alpha\tilde{g}(\omega)}\frac{1}{1+\mu^2g(-\omega)}\gamma(\omega)+\mathcal{O}(\lambda^2)
\ee
The neglected graphs contribute as $\mathcal{O}(\lambda^2)$. The constant $\alpha$ comes from the loop (Fig. \ref{fig_I_ord}) (the prefactor $1/2$ is the symmetry factor)
\be
\alpha=\frac{1}{2}\int_{-\infty}^\infty \frac{\dd \omega}{2\pi}:\gamma^\text{dr}(\omega):\, =\int_0^\infty \frac{\dd k}{2\pi}  \frac{2 E(k)v^2(k)}{4 v^2(k)+\mu^4} (\rho(k)+\rho(-k))
\ee
With the double dots, we mean the normal ordering must be considered.
Using now the relations \eqref{F_rel}, \eqref{A_rel} and \eqref{cd_rel} we can compute $\mathcal{A}_k$ and $\mathcal{F}_k$, and finally $\delta \rho(k)$ through Eq. \eqref{Sdeltarho}. This requires only simple algebraic manipulations and leads to the compact result
\be
\delta\rho(k)=\frac{(\mu^2+\lambda\alpha)^2}{4E^2(k)v^2(k)+(\mu^2+\lambda\alpha)^2}\Big[\rho(-k)-\rho(k)\Big]+\mathcal{O}(\lambda^2)
\ee
which has the same form of the non-interacting case (18), provided we replace $\mu^2\to\mu^2+\lambda\alpha$.
Incidentally, the expression of the first order in the quantum and classical case coincide.
Computing the next order in $\delta \rho$ is more cumbersome and involves several diagrams. For completeness, hereafter we report the result within the classical realm used in Fig. 1
\be\label{c_2_ord}
c(\omega)=\gamma_\text{dr}(\omega)+2\Re\left[\frac{\lambda^2}{2}\gamma_\text{dr}^{(1)}g_\text{dr}^{(1)}(\omega)\left(I_1(\omega)+I_0 I_3\right)\right]+\frac{\lambda^2}{6}I_2(\omega)|g_\text{dr}(\omega)|^2+\mathcal{O}(\lambda^3)
\ee
\be\label{d_2_ord}
d(\omega)=\left[1+\frac{\lambda^2}{2}g_\text{dr}^{(1)}(\omega)\left(I_1(\omega)+I_3I_0\right)\right]\frac{\gamma(\omega)}{1+(\mu^2+\alpha\lambda)g(\omega)}+\mathcal{O}(\lambda^3)
\ee

where the auxiliary functions are defined as:

\be
g_\text{dr}^{(1)}(\omega)=\frac{g(\omega)}{1+(\mu^2+\lambda\alpha)g(\omega)}\, ,\hspace{2pc} \gamma_\text{dr}^{(1)}(\omega)=\left|\frac{1}{1+(\mu^2+\lambda\alpha)g(\omega)}\right|^2\gamma(\omega)
\ee
\be
G^{(1)}_\text{dr}(t)=\int_{-\infty}^\infty \frac{\dd \omega}{2\pi} e^{i\omega t} g_{\text{dr}}^{(1)}(\omega)\,, \hspace{5pc}\Gamma^{(1)}_\text{dr}(t)=\int_{-\infty}^\infty \frac{\dd \omega}{2\pi} e^{i\omega t} \gamma_{\text{dr}}^{(1)}(\omega)
\ee
\be
I_0=\int_0^\infty \frac{\dd k}{2\pi} 2\Re\left[\gamma_\text{dr}^{(1)}(E(k))|v(k)|g_\text{dr}^{(1)}(E(k))\right]\, ,\hspace{2pc} I_1(\omega)=\int_{-\infty}^\infty \dd t e^{-i\omega t}\big[\Gamma_\text{dr}^{(1)}(t)\big]^2G_\text{dr}^{(1)}(t)
\ee

\be
I_2(\omega)=\int_{-\infty}^\infty \dd t e^{-i\omega t}\big[\Gamma_\text{dr}^{(1)}(t)\big]^3\, ,\hspace{3pc} I_3=\int_{-\infty}^\infty \frac{\dd \omega}{2\pi} \gamma_\text{dr}^{(1)}(\omega)
\ee

In order to obtain the result (\ref{c_2_ord}-\ref{d_2_ord}) we took advantage of the OPIs  when possible.

\section{The extended defect}
\label{sec_ext_def}

In this brief section we outline the necessary modifications in the case of an extended defect.
In this respect, we start directly from the generalization of Eq. \eqref{Sint_delta} to a segment

\be\label{Delta_inteq}
\hat{\phi}_t(x)=\hat{\psi}_t(x)- \mu^2\int_{-\Delta}^{\Delta}\dd y\int_0^\infty \dd\tau\,G_{t-\tau}(x-y) \hat{\phi}_\tau(y)-\lambda\int_{-\Delta}^{\Delta}\dd y\int_0^\infty \dd\tau\,G_{t-\tau}(x-y) :\delta V'(\hat{\phi}_\tau(y)):\, .
\ee

Similarly to what we did in the $\delta-$case, we define
\be\label{int_DeltaG}
\tilde{G}_t(x,x')=G_t(x-x')-\mu^2\int_{-\Delta}^{\Delta}\dd y\int_{-\infty}^\infty \dd\tau\, G_{t-\tau}(x-y)\tilde{G}_{\tau}(y,x')\,
\ee

\be
\hat{\varphi}_t(x)=\hat{\psi}_t(x)- \mu^2\int_{-\Delta}^{\Delta}\dd y\int_{0}^\infty \dd\tau\,G_{t-\tau}(x,y)\hat{\varphi}_\tau(y)
\ee

and recast Eq. \eqref{Delta_inteq} as
\be
\hat{\phi}_t(x)=\hat{\varphi}_t(x)-\lambda\int_{-\Delta}^{\Delta}\dd y\int_0^\infty \dd\tau\,\tilde{G}_{t-\tau}(x-y) :\delta V'(\hat{\phi}_\tau(y)):\, .
\ee

Hereafter, we generalize the proof of the convergence of the series presented in Section \ref{conv_cl}, in the classical realm and for the same class of interactions (i.e. satisfying  $|\delta V'(\phi)|\le C_1$ and $|\delta V'(\phi)-\delta V'(\phi')|\le C_2|\phi-\phi'|)$.
In principle, in the case of the $\phi^4$ interaction (either classical or quantum) we could retrace the passages of Section \ref{pert_ord} and show the finitness of any order of the expansion, however we will stick to the simpler content of Section \ref{conv_cl}.

It should appear clear that a straightforward generalization of the proof presented in Section \ref{conv_cl} can be obtained if we show
\be\label{s48}
 \int_0^\infty \dd t\, |\tilde{G}_t(x,y)|\le M<\infty\, \hspace{2pc} |x|, |y|\le \Delta\, .
\ee

As a matter of fact, if the above inequality holds true, with the same considerations of Eq. \eqref{conv_bound} we reach the bound 
\be
\max_{t,x}|\phi_t^{(n+1)}(x)-\phi_t^{(n)}(x)|\le \lambda 2\Delta M C_2  \max_{t,x}|\phi_t^{(n)}(x)-\phi_t^{(n-1)}(x)|\, .
\ee
Above, the maximum is taken over $t\in (0,\infty)$ and $x\in (-\Delta,\Delta)$. Repeating the passages of Eq. \eqref{ser_conv}, we reach an upperbound for the convergence radius $\lambda < 1/(2\Delta M C_2)$.

In order to show Eq. \eqref{s48}, it is useful to study the differential equation associated with the integral equation \eqref{int_DeltaG}, being $\tilde{G}_t(x,x')$ solution of
\be
\partial_t^2\tilde{G}_t(x,x')=\begin{cases}\partial_x^2 \tilde{G}_t(x,x')-(m^2+\mu^2)\tilde{G}_t(x,x') \hspace{2pc}&|x|\le \Delta\\ \,\\
\partial_x^2 \tilde{G}_t(x,x')-m^2\tilde{G}_t(x,x')\hspace{2pc}&|x|> \Delta
  \end{cases}
\ee

with the requirement of being causal $\tilde{G}_{t<0}(x,x')=0$ and the initial condition $\partial_t\tilde{G}_{t}(x,x')\big|_{t=0}=\delta(x-x')$. $\tilde{G}$ can be then written in terms of an orthonormal basis of the differential equation
\be
\tilde{G}_t(x,x')=\Theta(t)\int_0^\infty \dd k\,  \frac{\sin(E(k)t)}{E(k)}\Big(\theta_k^e(x)[\theta_k^e(x')]^*+\theta_k^o(x)[\theta_k^o(x')]^*\Big)\, ,
\ee
where $\theta^{e/o}_k(x)$ are even/odd in $x$ and are orthonormal

\be
\begin{cases}\label{s51}
\theta_k^e(x)=\mathcal{N}^e_k\cos\big(x\sqrt{k^2-\mu^2}\big) \hspace{4pc} &0<x\le\Delta\\
\theta_k^e(x)=\mathcal{N}^e_k \Big[a^e_k \cos(k(x-\Delta))+b_k^e\sin(k(x-\Delta))\Big]\hspace{4pc} &x>\Delta\\
\end{cases}
\ee
\be
\begin{cases}
\theta^o_k(x)=\mathcal{N}^o_k\sin\big(x\sqrt{k^2-\mu^2}\big) \hspace{4pc} &0<x\le\Delta\\
\theta_k^o(x)=\mathcal{N}^o_k \left[a^o_k \cos(k(x-\Delta))+b_k^o\sin(k(x-\Delta))\right]\hspace{4pc} &x>\Delta\\
\end{cases}
\ee
The coefficients are fixed by continuity of the function and of the first derivative at $x=\Delta$
\be
\begin{cases}a_k^e=\cos\big(\Delta\sqrt{k^2-\mu^2}\big)\\
b_k^e=-\frac{\sqrt{k^2-\mu^2}}{k}\sin\big(\Delta\sqrt{k^2-\mu^2}\big)
\end{cases}\, ,\hspace{3pc} \begin{cases}a_k^o=\sin\big( \Delta\sqrt{k^2-\mu^2}\big)\\
b_k^o=\frac{\sqrt{k^2-\mu^2}}{k}\cos\big(\Delta\sqrt{k^2-\mu^2}\big)
\end{cases}
\ee

A correct normalization requires
\be
|\mathcal{N}_k^{e/o}|^2=\frac{1}{\pi}\frac{1}{|a_k^{e/o}+ib_k^{e/o}|^2+|a_k^{e/o}-ib_k^{e/o}|^2}
\ee
which guarantees the orthonormality
\be
\int_{-\infty}^\infty \dd x\, \theta_k^e(x)[\theta^e_q(x)]^*=\int_{-\infty}^\infty \dd x\, \theta_k^o(x)[\theta^o_q(x)]^*=\delta(k-q)\, .
\ee

In order to ensure the validity of Eq. \eqref{s48}, we change integration variable $\omega=\sqrt{k^2+m^2}$ and perform an integration by parts
\begin{multline}
\int_m^\infty \dd \omega\,  \frac{\sin(\omega t)}{\sqrt{\omega^2-m^2}}\Big(\theta_k^e(x)[\theta_k^e(x')]^*+\theta_k^o(x)[\theta_k^o(x')]^*\Big)_{k=\sqrt{\omega^2-m^2}}=\\
\frac{1}{t}\left[ \frac{-\cos(\omega t)}{\sqrt{\omega^2-m^2}}\Big(\theta_k^e(x)[\theta_k^e(x')]^*+\theta_k^o(x)[\theta_k^o(x')]^*\Big)_{k=\sqrt{\omega^2-m^2}}\right]_{m}^{+\infty}+\\
\frac{1}{t}\int_m^\infty \dd \omega\,  \cos(\omega t)\partial_\omega\left[\frac{1}{\sqrt{\omega^2-m^2}}\Big(\theta_k^e(x)[\theta_k^e(x')]^*+\theta_k^o(x)[\theta_k^o(x')]^*\Big)_{k=\sqrt{\omega^2-m^2}}\right]
\end{multline}

The boundary term at $\omega\to+\infty$ clearly vanishes, for what it concerns the boundary term at $\omega=m$ it could appear singular at first sight, but it is not the case. We now look closely at the function
\be\label{S56}
\frac{1}{\sqrt{\omega^2-m^2}}\Big(\theta_k^e(x)[\theta_k^e(x')]^*+\theta_k^o(x)[\theta_k^o(x')]^*\Big)_{k=\sqrt{\omega^2-m^2}}
\ee
Notice that at $\omega\to m$, while the prefactor $(\omega^2-m^2)^{-1/2}$ diverges, the eigenfunctions $\theta_k^{e/o}$ are instead vanishing. In particular, assuming $x,x'$ both in $(-\Delta,\Delta)$, the magnitude of $\theta_k^{e/o}$ is ruled by $\mathcal{N}_k^{e/o}\propto k$ for $k\to 0$. Therefore, the boundary term at $\omega=m$ vanishes as $\sim \sqrt{\omega^2-m^2}$. So, the function \eqref{S56} is \emph{i)} continuous, \emph{ii)} vanishes at $\omega=m$  and \emph{iii)} has non analytic points at $\omega=m$ and $\omega=\sqrt{m^2+\mu^2}$. These are squareroot-singularities (similarly to Eq. \eqref{G_larget}) which allow for a large $t$ estimation
\be
\tilde{G}_t(x,x')\simeq t^{-3/2}\, ,\hspace{2pc} x,x'\in (-\Delta,\Delta)\, .
\ee
The above time decay is sufficient to ensure the validity of Eq. \eqref{s48}, since there are no singularities in the $x,x'$ variables.

\section{Galilean bosons and fermions}
\label{sec_gal}

Our considerations and diagrammatic approach can be easily extended to Galilean bosons and fermions. For example, we can choose
\be
H_\text{I}=\int \dd x \, \partial_x\hat{\chi}^\dagger(x)\partial_x\hat{\chi}(x)
\ee
where $\hat{\chi}(x)$ is either a quantum boson or fermion $[\hat{\chi}(x),\hat{\chi}^\dagger(y)]_{\pm}=\delta(x-y)$, or it can be interpreted as a classical field.
In this case a GGE constructed on the free model $H_\text{I}$ is simply gaussian and diagonal in the Fourier space of the $\hat{\chi}(x)$ fields
\be
\langle \hat{\chi}^\dagger(x)\hat{\chi}(y)\rangle_\text{GGE}=\int_{-\infty}^\infty \frac{\dd k}{2\pi} e^{ik(x-y)}\rho(k)\, .
\ee
For simplicity we consider a particle-conserving interaction in the following form
\be
\hat{V}(x)=\mu^2\hat{\chi}^\dagger(x)\hat{\chi}(x)+\lambda \delta V[\hat{\chi}^\dagger(x)\hat{\chi}(x)]\, ,
\ee
with $\delta V$ an analytic function of the density. However, generalizations which include derivatives of the fields as well as non local short-ranged interactions are straightforward. The repulsive potential $\mu^2\hat{\chi}^\dagger(x)\hat{\chi}(x)$, similarly to the relativistic case, is needed in order to prevent secular terms in the perturbative expansion.
The Heisenberg equation of motion for $\chi$ are then
\be
\partial_t\hat{\chi}_t(x)=\begin{cases}-\partial_x^2 \hat{\chi}_t(x)+\mu^2\chi_t(x)+\delta V'[\hat{\chi}_t^\dagger(x)\hat{\chi}_t(x)]\hat{\chi}_t(x)\hspace{2pc} & |x|<\Delta\\ \, \\ -\partial_x^2 \hat{\chi}_t(x)\hspace{2pc} & |x|>\Delta
\end{cases}
\ee

Equivalently, we can use the integral equation
\be\label{S63}
\hat{\chi}_t(x)=\hat{\psi}_t(x)+\int_0^\infty \dd \tau\,\int_{-\Delta}^\Delta \dd y\,  G_{t-\tau}(x-y) \Big[\mu^2\hat{\chi}_{\tau}(y)+\delta V'[\hat{\chi}_\tau^\dagger(y)\hat{\chi}_\tau(y)]\hat{\chi}_t(y)\Big]\, ,
\ee
where $\hat{\psi}_t(x)$ is the evolution of $\hat{\chi}$ in absence of the defect and the Green function being now
\be
G_t(x)=\Theta(t)\int_{-\infty}^\infty \frac{\dd k}{2\pi} e^{ikx -iE(k) t}\, , \hspace{2pc} E(k)=k^2\, .
\ee
From Eq. \eqref{S63}, all the steps described in the previous sections can be retraced: a perturbative expansion constructed starting from $\mu^2>0$ and $\lambda=0$ has the same convergence properties of the relativistic counterpart.

\section{Lattice models and self trapping}
\label{sec_lat}
As it has been mentioned in the main text, lattice models can display the phenomenon of self-trapping \ocite{Sself_trap} which plagues our perturbative approach.

The phenomenon is due to the fact that, on the lattice, the kinetic energy is bounded and, semiclassically, it can be explained as it follows: imagine we turn on a repulsive potential of strength $U>0$ in the defect region. A particle initially sat on the defect with momentum $k$, after the potential activation, will have a total energy $E(k)+U$, being $E(k)$ the kinetic energy.
Due to energy conservation, the particle can leave the defect region with a momentum $k'$ only if $E(k)+U=E(k')$: however, if the kinetic energy is bounded and $U$ is large enough there are not solutions to this equation. Therefore, despite the potential being repulsive, the particle is doomed to live within the defect region, thus forming a bound state.

In this section we point out these problems in a specific example, i.e. the lattice discretization of our relativistic model
\be
H_\text{I}=\sum_i \frac{1}{2}\hat{\Pi}^2(i)+\frac{1}{2}\big[\hat{\phi}(i+1)-\hat{\phi}(i)\big]^2+\frac{m^2}{2}\hat{\phi}^2(i)\, .
\ee
We consider for simplicity the natural discretization of the $\delta-$like defect, i.e. a potential acting non trivially only on a lattice site, which is placed at zero. The integral equation describing the solution to the Heisenberg equation of motion is (the lattice model does not have UV divergences and thus normal ordering is not necessary)
\be\label{int_dis}
\hat{\phi}_t(i)=\hat{\psi}_t(i)-\int_0^\infty\dd \tau\, G_{t-\tau}(i) V'(\hat{\phi}_{\tau}(0))\, .
\ee
The Green function in this case is
\be
G_t(x)=\Theta(t)\int_{-\pi}^\pi \frac{\dd k}{2\pi} \frac{e^{ikx}}{E(k)}\sin(E(k)t)\,, \hspace{2pc}E(k)=\sqrt{m^2+2(1-\cos k)}\, .
\ee

At large time, $G_t(0)\sim t^{-1/2}$: similarly to the continuum case, a perturbative expansion around $V=0$ is plagued with singularities, due to the modes with zero velocity (which are responsible of the $t^{-1/2}$ decay).
Following what we did in Section \ref{sec_delta}, we can add a repulsive potential on the defect and cure the divergence due to the zero-velocity modes, thus
\be
V(\hat{\phi})=\frac{\mu^2}{2}\hat{\phi}^2+\lambda \delta V(\hat{\phi})\, .
\ee
Similarly to Eq. \eqref{def_int}, Eq. \eqref{int_dis} on top of the defect $(i=0)$ can be rewritten as
\be
\hat{\phi}_t(0)=\hat{\varphi}_t-\lambda\int_0^\infty \dd\tau\,\tilde{G}_{t-\tau} :\delta V'(\hat{\phi}_\tau(0)):\, .
\ee

The Green function $\tilde{G}$ is easily written in terms of $G$ in the Fourier space, identically to Eq. \eqref{tilde_g_four}
\be\label{tilde_g_dis}
\tilde{g}(\omega)=\frac{g(\omega)}{1+\mu^2g(\omega)}\, ,
\ee
however, differently from the continuous case, $\tilde{g}$ is singular. In order to show this fact, consider $g(\omega)$
\be
g(\omega)=\int_{-\infty}^\infty \dd t\, e^{-i\omega t}G_t(0)=\int_{-\pi}^\pi \frac{\dd k}{2\pi} \left[\mathcal{P}\left(\frac{1}{E^2(k)-\omega^2}\right)-i\text{sign}(\omega)\delta(E^2(k)-\omega^2)\right]\, ,
\ee
where the principal part prescription is used to handle the pole singularities.
Notice that the above integral is singular when $\omega=\pm E(k)$ with $k$ such that $\partial_k E(k)=v(k)=0$: these singularities are cured in $\tilde{g}(\omega)$ \eqref{tilde_g_dis} exactly as it happened in the continuum case. However, if $E(k)$ is bounded (as it happens in lattice models) $\tilde{g}(\omega)$ acquires a new singularity.
Indeed, for $|\omega|>\max_k |E(k)|$, $g(\omega)$ is purely real and negative, furthermore $\lim_{\omega\to (\max_k |E(k)|)^+}g(\omega)=-\infty$ and $\lim_{\omega\to \infty}g(\omega)=0$.
This necessary implies the existence of $\bar{\omega}$ such that $1+\mu^2g(\bar{\omega})=0$, which ultimately is translated in a pole-singularity for $\tilde{g}$ \eqref{tilde_g_four}.

Poles in the frequency space of Green functions are known to be associated with bound states, which plague the perturbative expansion: semiclassically, a particle is trapped in the defect region and keeps interacting with the others, eventually building up a non-perturbative effect.

\section{Numerical method}
\label{sec_num}

This section is dedicated to describe the numerical method used in Fig. 1. The classical theory is discretized
\be
\phi(x)\to \phi_j\hspace{3pc}\Pi(x)\to \Pi_j\, ,
\ee
a lattice space $\ell$ is introduced and the system is put in finite size $L=N\ell$, with $N$ the number of lattice sites. We impose periodic boundary conditions.
The evolution of the expectation value of a given observable $\langle \mathcal{O}_t\rangle$ is computed as it follows
\begin{enumerate}
\item A random initial condition $\phi_j$, $\Pi_j$ is chosen. The probability distribution is determined by the pre-quench GGE.
\item The initial condition is deterministically evolved and $\mathcal{O}_t$ computed for a single realization.
\item The previous steps are repeated for several realizations and $\langle \mathcal{O}_t\rangle$ is computed averaging on the initial conditions.
\end{enumerate}

The main difficulty resides in a good approximation of the continuum model by mean of a lattice regularization. In principle, we could have discretized the problem in the real space similarly to Section \ref{sec_lat} and then numerically solve the discrete model. However, after a first analysis, this approach seemed to lead to large errors caused by the finite lattice space, therefore we opted for a different method.

For what it concerns the initial conditions, we use the mode expansion
\be\label{mode_dec}
\phi_j=\frac{1}{\sqrt{N\ell}}\sum_{n=0}^{N-1} \frac{\sqrt{\rho_{N,\ell}(n)}}{\sqrt{2 E_{N,\ell}(n)}}\Big(e^{-ijn2\pi/N}\eta_n+\text{c.c.}\Big)\hspace{2pc} \Pi_j=\frac{1}{\sqrt{N\ell}}\sum_{n=0}^{N-1} \sqrt{\frac{E_{N,\ell}(n)\rho_{N,\ell}(n)}{2}}\Big(-ie^{-i2\pi jn/N}\eta_n+\text{c.c.}\Big)\,
\ee
where c.c. stands for the complex conjugated and
\be
E_{N,\ell}(n)=\begin{cases}\sqrt{m^2+[2\pi /(\ell N)]^2 n^2} \hspace{2pc} &n<N/2\\
\, \\ \sqrt{m^2+[2\pi /(\ell N)]^2 (N-n)^2} \hspace{2pc} &n\ge N/2
\end{cases}\, ,\hspace{2pc}\rho_{N,\ell}(n)=\begin{cases}\rho\big(2\pi n/(\ell N)\big) \hspace{2pc} &n<N/2\\
\, \\ \rho\big(2\pi(N-n)/(\ell N)\big) \hspace{2pc} &n\ge N/2
\end{cases}
\ee

Above $\rho(k)$ is the root density of the GGE in the continuous model.
The modes $\eta_n$ are complex uncorrelated gaussian variables with variance $\langle |\eta_n|^2\rangle=1$, the probability distribution of the phase is flat.
It is a simple exercise to compute the two point correlator and recover the continuum GGE-expectation values in the limit $N\to\infty$ and $\ell\to0$
\be
\langle \phi_j\phi_{j'}\rangle_\text{GGE}\xrightarrow{N\to\infty\,, \ell\to 0}\langle \phi(\ell j)\phi(\ell j')\rangle_\text{GGE}\, .
\ee 

For what it concerns the time evolution, we used a ``Trotterization" procedure discussed hereafter. 
We describe the algorithm in the case of a $\delta-$like defect which, in the discretized version, will act non trivially only on a lattice site which we choose to be $j=0$.
We assume $V'(\phi)$ can be written as
\be
V'(\phi)=U^2(\phi)\phi\hspace{2pc} U(\phi)>0\,
\ee
which is surely the case for $V(\phi)=\mu^2\phi^2/2+\lambda\phi^4/4!$, i.e. the potential we used as example.
We discretize the time evolution in steps $\Delta t$ and the method is accurate up to $\mathcal{O}(\Delta t^2)$.

The infinitesimal solution is carried out in two steps: in the first step we evolve the field as if the defect was absent, then we consider the effect of the impurity.
Given the field configuration $\phi_j(t)$ $\Pi_j(t)$ we construct the mode expansion similarly to  \eqref{mode_dec}
\be
\phi_{j,t}=\frac{1}{\sqrt{N\ell}}\sum_{n=0}^{N-1} \frac{\sqrt{\rho_{N,\ell}(n)}}{\sqrt{2 E_{N,\ell}(n)}}\Big(e^{-ijn2\pi/N}\eta_n(t)+\text{c.c.}\Big)\hspace{2pc} \Pi_{j,t}=\frac{1}{\sqrt{N\ell}}\sum_{n=0}^{N-1} \sqrt{\frac{E_{N,\ell}(n)\rho_{N,\ell}(n)}{2}}\Big(-ie^{-i2\pi jn/N}\eta_n(t)+\text{c.c.}\Big)\,
\ee
the modes $\eta_j(t)$ are \emph{defined} by the above equation. Then, an intermediate field configuration $\phi'_{j,t}$, $\Pi'_{j,t}$ is constructed evolving the modes $\eta_n(t)\to e^{-i\Delta t E_{n,\ell}(j)}\eta_n(t)$
\be
\phi_{j',t}=\frac{1}{\sqrt{N\ell}}\sum_{n=0}^{N-1} \frac{\sqrt{\rho_{N,\ell}(n)}}{\sqrt{2 E_{N,\ell}(n)}}\Big(e^{-ijn2\pi/N-i\Delta t E_{N,\ell}(n)}\eta_n(t)+\text{c.c.}\Big)
\ee
and similarly for $\Pi'_{j,t}$. Then, we act with the defect: for $j\ne0$ we set $\phi_{j,t+\Delta t}=\phi'_{j,t}$ and $\Pi_{j,t+\Delta t}=\Pi'_{j,t}$, while for $j=0$

\be
\Pi_{0,t+\Delta t}=-i\sqrt{\frac{ U(\phi'_{0,t})}{2}}(z-z^*)\hspace{2pc}\phi_{0,t+\Delta t}=\frac{1}{\sqrt{2 U(\phi'_{0,t})}}(z+z^*)
\ee
where

\be
z=e^{-i\Delta t U(\phi'_{0,t})}\sqrt{\frac{U(\phi'_{0,t})}{2}}\left(\phi'_{0,t}+\frac{i}{U(\phi'_{0,t})}\Pi'_{0,t}\right)
\ee

Going back and forth from the real and Fourier space is surely computationally expensive, however for what it concerns the time evolution the only information we need is the fields at $j=0$, which can be computed with $\sim N$ steps from the knowledge of the Fourier modes. Therefore, implementing the algorithm in the Fourier space the time evolution can be carried up to time $t_\text{max}$ in $\sim N t_\text{max}/\Delta t$ steps.
The most time consuming operation is evaluating observables in real space for which the whole $\phi_{i,t}$ is needed: using a fast Fourier transform algorithm, we can pass from the Fourier space to the real one in $N\log N$ steps.


\begin{thebibliography}{99}

\bibitem{exp1}
M. Greiner \emph{et al}, Nature \href{\doi10.1038/nature00968}{\bf 419}, 51-54 (2002).

\bibitem{exp2}
T. Kinoshita, T. Wenger,  and D. S. Weiss,  Nature \href{\doi10.1038/nature04693}{\bf 440}, 900 (2006). 

\bibitem{exp3}
S. Hofferberth, I. Lesanovsky \emph{et al}, Nature \href{\doi10.1038/nature06149}{\bf 449}, 324-327 (2007). 

\bibitem{exp4}
L. Hackermuller, U. Schneider \emph{et al}, Science \href{\doi10.1126/science.1184565}{\bf 327}, 1621 (2010).

\bibitem{exp5}
S. Trotzky, Y.-A. Chen \emph{et al}, Nature Phys. \href{\doi10.1038/nphys2232}{\bf 8}, 325 (2012).

\bibitem{exp6}
M. Gring, M. Kuhnert \emph{et al}, Science \href{\doi10.1126/science.1224953}{\bf 337}, 1318 (2012).

\bibitem{exp7}
M. Cheneau, P. Barmettler \emph{et al}, Nature \href{http://dx.doi.org/10.1038/nature10748}{\bf 481}, 484 (2012).

\bibitem{exp8}
T. Langen, R. Geiger \emph{et al}, Nature Physics \href{http://dx.doi.org/10.1038/nphys2739}{\bf 9}, 640 (2013).

\bibitem{exp9}
F. Meinert, M.J. Mark \emph{et al}, Phys. Rev. Lett. \href{http://dx.doi.org/10.1103/PhysRevLett.111.053003}{\bf 111}, 053003 (2013).

\bibitem{exp10}
J.P. Ronzheimer, M. Schreiber \emph{et al}, Phys. Rev. Lett. \href{\doi10.1103/PhysRevLett.110.205301}{\bf 110}, 205301 (2013).

\bibitem{exp11}
 L. Vidmar, J.P. Ronzheimer, M. Schreiber, S. Braun, S.S. Hodgman, S. Langer, F. Heidrich-Meisner, I. Bloch, and U. SchneiderPhys. Rev. Lett. \href{\doi10.1103/PhysRevLett.115.175301}{\bf 115}, 175301 (2015).

\bibitem{exp12}
I. Bloch, Nature Physics \href{\doi 10.1038/nphys138}{\bf 1}, 23-30 (2005).

\bibitem{exp13}
 B. Paredes, Q. Widera, V. Murg, O. Mandel, S. F\"olling, I. Cirac, G. V. Shlyapnikov, T. W. H\"ansch, I. Bloch, Nature \href{\doi 10.1038/nature02530}{\bf 429}, 277-281 (2004)

\bibitem{exp14}
 R. J\"ordens, N. Strohmaier, K. G\"unter, H. Moritz, T. Esslinger, Nature \href{\doi 10.1038/nature07244}{\bf 455}, 204-207 (2008).

\bibitem{exp15}
S. Palzer, C. Zipkes, C. Sias, M. K\"ohl, Phys. Rev. Lett. \href{https://doi.org/10.1103/PhysRevLett.103.150601}{\bf 103}, 150601 (2009).


\bibitem{Korepin}
 V.E. Korepin, N.M. Bogoliubov, A.G. Izergin, \emph{Quantum Inverse Scattering Method and Correlation Functions} (University Press, Cambridge, 1993).

\bibitem{smirnov}
  F. A. Smirnov, \emph{Form factors in completely integrable models of quantum field theory }(World Scientific, 1992).

\bibitem{takahashi}
 M. Takahashi, \emph{Thermodynamics of one-dimensional solvable models} (Cambridge University Press, 2005).

\bibitem{calabrese-cardy}
P. Calabrese and  J. Cardy, Phys. Rev. Lett. \href{\doi10.1103/PhysRevLett.96.136801}{\bf 96}, 136801 (2006);  J. Stat. Mech. (2007) \href{\doi10.1088/1742-5468/2007/06/P06008}{P06008}.

\bibitem{ggew1}
F.H.L. Essler and M. Fagotti, J. Stat. Mech. (2016) \href{\doi10.1088/1742-5468/2016/06/064002}{064002}.

\bibitem{ggew2}
J.-S. Caux and F.H.L. Essler, Phys. Rev. Lett. \href{http://dx.doi.org/10.1103/PhysRevLett.110.257203}{\bf 110}, 257203 (2013).

\bibitem{ggew3}
B. Pozsgay, J. Stat. Mech. (2014) \href{\doi10.1088/1742-5468/2014/10/P10045}{P10045} .

\bibitem{ggew4}
P. Calabrese, F. H. L. Essler and M. Fagotti,  Phys. Rev. Lett. \href{http://dx.doi.org/10.1103/PhysRevLett.106.227203}{\bf 106}, 227203 (2011).

\bibitem{ggew5}
M. Fagotti and F.H.L. Essler, Phys. Rev. B \href{http://dx.doi.org/10.1103/PhysRevB.87.245107}{\bf 87}, 245107 (2013).

\bibitem{ggew6}
F. H. L. Essler, S. Evangelisti, and M. Fagotti, Phys. Rev. Lett.  \href{http://dx.doi.org/10.1103/PhysRevLett.109.247206}{\bf 109}, 247206 (2012).

\bibitem{ggew7}
 B. Pozsgay, J. Stat. Mech. (2013) \href{http://dx.doi.org/10.1088/1742-5468/2013/07/P07003}{P07003}.

\bibitem{ggew8}
M. Fagotti and F. H. L. Essler, J. Stat. Mech. (2013) \href{http://dx.doi.org/10.1088/1742-5468/2013/07/P07012}{P07012}.

\bibitem{ggew9}
 J.-S. Caux and R. M. Konik, Phys. Rev. Lett. \href{http://dx.doi.org/10.1103/PhysRevLett.109.175301}{\bf 109}, 175301 (2012).

\bibitem{ggew10}
 G. Mussardo, Phys. Rev. Lett. \href{http://dx.doi.org/10.1103/PhysRevLett.111.100401} {\bf 111}, 100401 (2013).

\bibitem{specialissue}
 Special issue on {\em Quantum Integrability in Out of Equilibrium Systems}, editors P. Calabrese, F.H.L. Essler and G. Mussardo, J. Stat. Mech. (2016) 064001.

\bibitem{ViRi2016}L. Vidmar and M. Rigol J. Stat. Mech. (2016) \href{https://doi.org/10.1088/1742-5468/2016/06/064007}{064007}.

\bibitem{ggerigol}
 M. Rigol, V. Dunjko, V. Yurovsky, and M. Olshanii,Phys. Rev. Lett. \href{http://dx.doi.org/10.1103/PhysRevLett.98.050405}{\bf 98}, 050405  (2007).

\bibitem{ggef1}
B. Pozsgay, M. Mesty\'{a}n , M. A. Werner, M. Kormos, G. Zar\'{a}nd, and G. Tak\'{a}cs, Phys. Rev. Lett. \href{http://dx.doi.org/10.1103/PhysRevLett.113.117203}{{\bf 113}}, 117203 (2014).

\bibitem{ggef2}
M. Mestyan, B. Pozsgay, G. T\'akacs, M.A. Werner, J. Stat. Mech. (2015) \href{\doi10.1088/1742-5468/2015/04/P04001}{P04001}.

\bibitem{ggef3}
J. De Nardis, B. Wouters, M. Brockmann, and J.-S. Caux, Phys. Rev. A \href{http://dx.doi.org/10.1103/PhysRevA.89.033601}{\bf 89}, 033601 (2014).

\bibitem{ggef4}
B. Wouters, J. De Nardis, M. Brockmann, D. Fioretto, M. Rigol, and J.-S. Caux, Phys. Rev. Lett.  \href{http://dx.doi.org/10.1103/PhysRevLett.113.117202}{\bf 113}, 117202 (2014).

\bibitem{lch1}
 E. Ilievski, J. De Nardis, B. Wouters, J-S Caux, F. H. L. Essler, T. Prosen, Phys. Rev. Lett. \href{\doi10.1103/PhysRevLett.115.157201}{\bf 115}, 157201 (2015). 

\bibitem{lch2}
E. Ilievski, M. Medenjak, and T. Prosen, Phys. Rev. Lett. \href{http://dx.doi.org/10.1103/PhysRevLett.115.120601}{\bf 115}, 120601 (2015).

\bibitem{lch3}
E. Ilievski, M. Medenjak \emph{et al}, J. Stat. Mech. (2016) \href{\doi10.1088/1742-5468/2016/06/064008}{064008}.

\bibitem{lch4}
E. Ilievski, E. Quinn, J. De Nardis, M. Brockmann J. Stat. Mech. (2016) \href{\doi10.1088/1742-5468/2016/06/063101}{P063101}.

\bibitem{lch5}
L. Piroli, E. Vernier, J. Stat. Mech. (2016) \href{\doi10.1088/1742-5468/2016/05/053106}{P053106}.

\bibitem{lch6}
 L. Piroli, E. Vernier, P. Calabrese, Phys. Rev. B \href{\doi10.1103/PhysRevB.94.054313}{\bf 94}, 054313 (2016).

\bibitem{lchf1}
S. Sotiriadis, Phys. Rev. A \href{https://doi.org/10.1103/PhysRevA.94.031605}{\bf 94}, 031605 (2016).

\bibitem{lchf2}
F. H. L. Essler, G. Mussardo, and M. Panfil,  Phys. Rev. A \href{http://dx.doi.org/10.1103/PhysRevA.91.051602}{\bf 91}, 051602(R) (2015).

\bibitem{lchf3}
A. Bastianello, S. Sotiriadis, J. Stat. Mech \href{\doi10.1088/1742-5468/aa5738 }{023105} (2017).

\bibitem{lchf4}
 F. H. L. Essler, G. Mussardo, M. Panfil, J. Stat. Mech. (2017) \href{http://stacks.iop.org/1742-5468/2017/i=1/a=013103}{013103}.

\bibitem{lchf5}
 E. Vernier and A. C. Cubero, J. Stat. Mech. (2017) \href{http://stacks.iop.org/1742-5468/2017/i=2/a=023101}{023101}.

\bibitem{bertini_prethermal}B. Bertini, F. H. L. Essler, S. Groha, and Neil J. Robinson, Phys. Rev. Lett. \href{https://doi.org/10.1103/PhysRevLett.115.180601}{\bf115}, 180601 (2015); Phys. Rev. B \href{https://doi.org/10.1103/PhysRevB.94.245117}{\bf94}, 245117 (2016).



\bibitem{pre_th_1}M. Moeckel and S. Kehrein, Phys. Rev. Lett. \href{https://doi.org/10.1103/PhysRevLett.100.175702}{\bf100}, 175702
(2008); Ann. Phys. (Amsterdam) \href{https://doi.org/10.1016/j.aop.2009.03.009}{\bf324}, 2146 (2009).

\bibitem{pre_th_2} A. Rosch, D. Rasch, B. Binz, and M. Vojta, Phys. Rev. Lett.
\href{https://doi.org/10.1103/PhysRevLett.101.265301}{\bf101}, 265301 (2008).

\bibitem{pre_th_3} M. Kollar, F. A. Wolf, and M. Eckstein, Phys. Rev. B \href{https://doi.org/10.1103/PhysRevB.84.054304}{\bf84},
054304 (2011).

\bibitem{pre_th_4} M. van den Worm, B. C. Sawyer, J. J. Bollinger, and M.
Kastner, New J. Phys. \href{https://doi.org/10.1088/1367-2630/15/8/083007}{\bf15}, 083007 (2013).

\bibitem{pre_th_5} M. Marcuzzi, J. Marino, A. Gambassi, and A. Silva, Phys.
Rev. Lett. \href{https://doi.org/10.1103/PhysRevLett.111.197203}{\bf111}, 197203 (2013).

\bibitem{pre_th_6} F. H. L. Essler, S. Kehrein, S. R. Manmana, and N. J.
Robinson, Phys. Rev. B \href{https://doi.org/10.1103/PhysRevB.89.165104}{\bf89}, 165104 (2014).

\bibitem{pre_th_7} N. Nessi, A. Iucci, and M. A. Cazalilla, Phys. Rev. Lett. \href{https://doi.org/10.1103/PhysRevLett.113.210402}{\bf113}, 210402 (2014).

\bibitem{pre_th_8}M. Fagotti, J. Stat. Mech. (2014) \href{https://doi.org/10.1088/1742-5468/2014/03/P03016}{P03016}.

\bibitem{pre_th_9} G. P. Brandino, J.-S. Caux, and R. M. Konik
Phys. Rev. X \href{https://doi.org/10.1103/PhysRevX.5.041043}{\bf5}, 041043 (2015).

\bibitem{pre_th_10} B. Bertini and M. Fagotti, J. Stat. Mech. (2015) \href{https://doi.org/10.1088/1742-5468/2015/07/P07012}{P07012}.

\bibitem{pre_th_11}  C. Kollath, A. M. Läuchli, and E. Altman, Phys. Rev. Lett.
\href{https://doi.org/10.1103/PhysRevLett.98.180601}{\bf98}, 180601 (2007)

\bibitem{pre_th_12}M. Rigol, Phys. Rev. Lett. \href{https://doi.org/10.1103/PhysRevLett.103.100403}{\bf103}, 100403 (2009).

\bibitem{pre_th_13}A. Mitra, Phys. Rev. B \href{https://doi.org/10.1103/PhysRevB.87.205109}{\bf87}, 205109 (2013).

\bibitem{pre_th_14} A. Chiocchetta, M. Tavora, A. Gambassi, and A. Mitra
Phys. Rev. B \href{https://doi.org/10.1103/PhysRevB.91.220302}{\bf91}, 220302(R) (2015).

\bibitem{pre_th_15}A. Chiocchetta, A. Gambassi, S. Diehl, and J. Marino, Phys. Rev. Lett.
\href{https://doi.org/10.1103/PhysRevLett.118.135701}{\bf118}, 135701 (2017).


\bibitem{Alba_Fagotti_prethermal} V. Alba, M. Fagotti, Phys. Rev. Lett. \href{https://doi.org/10.1103/PhysRevLett.119.010601}{\bf119}, 010601 (2017).


\bibitem{exp_pre_th1} M. Gring, M. Kuhnert, T. Langen, T. Kitagawa, B. Rauer,
M. Schreitl, I. Mazets, D. Adu Smith, E. Demler, and J.
Schmiedmayer, Science \href{\doi10.1126/science.1224953}{\bf337}, 1318 (2012).

\bibitem{exp_pre_th2} D. Adu Smith, M. Gring, T. Langen, M. Kuhnert, B. Rauer,
R. Geiger, T. Kitagawa, I. Mazets, E. Demler, and J.
Schmiedmayer, New J. Phys. \href{https://doi.org/10.1088/1367-2630/15/7/075011}{\bf15}, 075011 (2013).

\bibitem{exp_pre_th3} T. Langen, M. Gring, M. Kuhnert, B. Rauer, R. Geiger,
D. A. Smith, I. E. Mazets, and J. Schmiedmayer, Eur. Phys.
J. Spec. Top. \href{https://doi.org/10.1140/epjst/e2013-01752-0}{\bf217}, 43 (2013).

\bibitem{exp_pre_th4} Y. Tang, W. Kao, K.-Y. Li, S. Seo, K. Mallayya, M. Rigol, S. Gopalakrishnan, and B. L. Lev, Phys. Rev. X \href{https://doi.org/10.1103/PhysRevX.8.021030}{\bf8}, 021030 (2018).

\bibitem{localquenchesBertini_Fagotti}
B. Bertini and M. Fagotti, Phys. Rev. Lett. \href{\doi10.1103/PhysRevLett.117.130402}{\bf 117} 130402 (2016).


\bibitem{fag_non_th_def} M. Fagotti, J. Phys. A: Math. Theor. \href{https://doi.org/10.1088/1751-8121/50/3/034005}{\bf50} 034005 (2017).




\bibitem{GHD1}
 O. A. Castro-Alvaredo, B. Doyon, T. Yoshimura, Phys. Rev. X \href{\doi10.1103/PhysRevX.6.041065}{\bf 6}, 041065 (2016).

\bibitem{GHD2}
B. Bertini, M. Collura, J. De Nardis, M. Fagotti, Phys. Rev. Lett. \href{\doi10.1103/PhysRevLett.117.207201}{\bf 117}, 207201 (2016).



\bibitem{DoyonSphon17} B. Doyon and H. Spohn, \href{http://dx.doi.org/10.21468/SciPostPhys.3.6.039}{SciPost Phys. 3, 039} (2017).

\bibitem{Doyon17} B. Doyon, \href{\doi 10.21468/SciPostPhys.5.5.054}{SciPost Phys. 5, 054 (2018).}


\bibitem{GHD3}
 B. Doyon, T. Yoshimura, \href{\doi10.21468/SciPostPhys.2.2.014}{SciPost Phys. 2, 014} (2017).

\bibitem{GHD6}
 B. Doyon, T. Yoshimura, J.-S. Caux, Phys. Rev. Lett. \href{10.1103/PhysRevLett.120.045301}{\bf120}, 045301 (2018).

\bibitem{GHD7}
 V. B. Bulchandani, R. Vasseur, C. Karrasch, J. E. Moore,Phys. Rev. Lett. \href{\doi 10.1103/PhysRevLett.119.220604}{\bf 119}, 220604 (2017),

\bibitem{GHD8}
 V. B. Bulchandani, R. Vasseur, C. Karrasch, J. E. Moore,Phys. Rev. B \href{\doi10.1103/PhysRevB.97.045407}{\bf 97}, 045407 (2018).



\bibitem{GHD10}
L. Piroli, J. De Nardis, M. Collura, B. Bertini, M. Fagotti, Phys. Rev. B \href{\doi 10.1103/PhysRevB.96.115124}{\bf 96}, 115124 (2017).


\bibitem{F17}
 M. Fagotti, 	Phys. Rev. B \href{\doi10.1103/PhysRevB.96.220302}{\bf 96}, 220302 (2017).

\bibitem{DS}
 B. Doyon and H. Spohn, J. Stat. Mech. \href{http://iopscience.iop.org/article/10.1088/1742-5468/aa7abf/meta}{073210} (2017).

\bibitem{ID117}
 E. Ilievski and J. De Nardis, Phys. Rev. Lett. \href{https://journals.aps.org/prl/abstract/10.1103/PhysRevLett.119.020602}{\bf 119}, 020602 (2017).

\bibitem{DDKY17}
 B. Doyon, J. Dubail, R. Konik and T. Yoshimura, Phys. Rev. Lett. \href{https://doi.org/10.1103/PhysRevLett.119.195301}{\bf 119}, 195301 (2017).

\bibitem{DSY17}
 B. Doyon,  H. Spohn and T. Yoshimura, Nucl. Phys. B  \href{https://doi.org/10.1016/j.nuclphysb.2017.12.002}{\bf 926}, 570-583 (2018).


\bibitem{ID217}
 E. Ilievski, J. De Nardis, Phys. Rev. B \href{https://journals.aps.org/prb/abstract/10.1103/PhysRevB.96.081118}{\bf 96}, 081118 (2017).

\bibitem{CDV17}
 M. Collura, A. De Luca, J. Viti, Phys. Rev. B \href{\doi 10.1103/PhysRevB.97.081111}{\bf 97}, 081111 (2018).




\bibitem{mazza2018}
L. Mazza, J. Viti, M. Carrega, D. Rossini, and Andrea De Luca, 	Phys. Rev. B \href{\doi10.1103/PhysRevB.98.075421}{\bf 98}, 075421 (2018).

\bibitem{BFPC18}
B. Bertini, M. Fagotti, L. Piroli, P. Calabrese,	J. Phys. A: Math. Theor. \href{\doi	10.1088/1751-8121/aad82e}{\bf 51}, 39LT01 (2018).

\bibitem{localquenches4}
M. Ljubotina, S. Sotiriadis, T. Prosen, \href{https://arxiv.org/abs/1802.05697}{arXiv:1802.05697} (2018).

\bibitem{Bas_DeL_hop} A. Bastianello, A. De Luca, Phys. Rev. Lett. \href{\doi10.1103/PhysRevLett.120.060602}{\bf120}, 060602 (2018).
\bibitem{Bas_DeL_ising} A. Bastianello, A. De Luca, Phys. Rev. B \href{https://doi.org/10.1103/PhysRevB.98.064304}{98}, 064304 (2018).




\bibitem{bedo15}D. Bernard, B. Doyon, Ann. Henri Poincaré (2015) \href{https://doi.org/10.1007/s00023-014-0314-8}{\bf16}: 113. 

\bibitem{fag15} M. Fagotti,  \href{https://arxiv.org/abs/1508.04401}{arXiv:1508.04401} (2015).

\bibitem{ruelle2000} D. Ruelle, J. of Statistical Physics (2000) \href{ https://doi.org/10.1023/A:1018618704438}{\bf 98}: 57.

\bibitem{GoMiOl18} J. P. Gomez, A. Minguzzi, M. Olshanii, \href{https://arxiv.org/abs/1806.01820}{arXiv:1806.01820} (2018).

\bibitem{PoAhHe18} J. Polo, V. Ahufinger, F. W. J. Hekking, and A. Minguzzi, Phys. Rev. Lett. \href{https://doi.org/10.1103/PhysRevLett.121.090404}{\bf121}, 090404 (2018).




\bibitem{localquenches1}
A. De Luca, Phys. Rev. B \href{\doi10.1103/PhysRevB.90.081403}{\bf 90} 081403 (2014).

\bibitem{localquenches3}
M. Fagotti, \href{https://arxiv.org/abs/1508.04401}{arXiv:1508.04401} (2015).







\bibitem{dmrg_rev}U. Schollwöck, Rev. Mod. Phys. \href{https://doi.org/10.1103/RevModPhys.77.259}{\bf77}, 259 (2005).


\bibitem{DeLuca_Mussardo2016} A. De Luca, G. Mussardo, J. Stat. Mech. (2016) \href{https://doi.org/10.1088/1742-5468/2016/06/064011}{064011}.
\bibitem{BDWY2018} A. Bastianello, B. Doyon, G. Watts, T. Yoshimura, \href{\doi10.21468/SciPostPhys.4.6.045}{SciPost Phys. 4, 045} (2018).


\bibitem{sakurai} J. J. Sakurai, E. D.  Commins, \emph{ Modern quantum mechanics, revised edition} (1995).

\bibitem{Nessi_Iucci}N Nessi and A Iucci 2014 J. Phys.: Conf. Ser. \href{https://doi.org/10.1088/1742-6596/568/1/012013}{\bf568} 012013.

\bibitem{YuOl10} V. A. Yurovsky, M. Olshanii, Phys. Rev. A \href{https://doi.org/10.1103/PhysRevA.81.043641}{\bf 81}, 043641 (2010).

\bibitem{suppl}Supplementary Material at url...


\bibitem{peskin} M. E. Peskin and D. V. Schroeder, \emph{An Introduction to Quantum
Field Theory} (Perseus, Reading, MA, 1995).


\bibitem{gaussification1}S. Sotiriadis, P. Calabrese, J. Stat. Mech. (2014) \href{http://dx.doi.org/10.1088/1742-5468/2014/07/P07024}{P07024}.
\bibitem{gaussification2}S. Sotiriadis, G. Martelloni, J. Phys. A \href{\doi10.1088/1751-8113/49/9/095002}{\bf4 9} 095002 (2016).
\bibitem{gaussification3}M. Gluza, C. Krumnow, M. Friesdorf, C. Gogolin, J. Eisert, Phys. Rev. Lett. \href{\doi10.1103/PhysRevLett.117.190602}{\bf 117} 190602 (2016).
\bibitem{gaussification4}S. Sotiriadis, 
J. Phys. A: Math. Theor. \href{https://doi.org/10.1088/1751-8121/aa8aa5}{\bf50} 424004 (2017).

\bibitem{self_trap} F. Bloch, Z. Phys. 52, 555 (1928); C. Zener, Proc.R. Soc. London A \href{\doi10.1098/rspa.1934.0116 }{\bf145}, 523 (1934).





\end{thebibliography}

\begin{thebibliography}{99}

\bibitem{Speskin} M. E. Peskin and D. V. Schroeder, \emph{An Introduction to Quantum
Field Theory} (Perseus, Reading, MA, 1995).

\bibitem{Sgaussification1}S. Sotiriadis, P. Calabrese, J. Stat. Mech. (2014) \href{http://dx.doi.org/10.1088/1742-5468/2014/07/P07024}{P07024}.
\bibitem{Sself_trap} F. Bloch, Z. Phys. 52, 555 (1928); C. Zener, Proc.R. Soc. London A \href{\doi10.1098/rspa.1934.0116 }{\bf145}, 523 (1934).


\end{thebibliography}
\end{document}